\documentclass[preprint,12pt,authoryear]{elsarticle}

\usepackage{lineno,hyperref}
\modulolinenumbers[5]

\usepackage{amssymb}
\usepackage{natbib,txfonts}
\usepackage{longtable}
\usepackage{graphics,epsfig,float}
\usepackage{pdflscape}

\journal {New Astronomy}

\DeclareRobustCommand{\ion}[2]{%
\relax\ifmmode
\ifx\testbx\f@series
{\mathbf{#1\,\mathsc{#2}}}\else
{\mathrm{#1\,\mathsc{#2}}}\fi
\else\textup{#1\,{\mdseries\textsc{#2}}}%
\fi}

\begin{document}
\begin{frontmatter}

\defcitealias{k13}{K13}
\defcitealias{zejda}{Z12}

\title{New variable stars discovered in the fields of three Galactic open clusters using the VVV Survey}


\author[mymainaddress,mysecondaryaddress,mythirdaddress]{T. Palma\corref{mycorrespondingauthor}}
\cortext[mycorrespondingauthor]{Corresponding author}
\ead{astrofisica.tp@unab.cl}

\author[mysecondaryaddress,mymainaddress]{D. Minniti}
\ead{dante@astrofisica.cl}

\author[mymainaddress,myforthaddress]{I. D\'ek\'any}
\ead{idekany@astro.puc.cl}

\author[mythirdaddress,myfifthaddress]{J.J. Clari\'a}
\ead{claria@oac.unc.edu.ar}

\author[mysixthaddress,mymainaddress]{J. Alonso-Garc\'ia}
\ead{javier.alonso@uantof.cl}

\author[mythirdaddress]{L.V. Gramajo}
\ead{luciana@oac.unc.edu.ar}

\author[mymainaddress,myseventhaddress]{S. Ram\'irez Alegr\'ia}
\ead{sebastian.ramirez@uv.cl}

\author[myeighthaddress]{C. Bonatto}
\ead{charles.bonatto@ufrgs.br}

\address[mymainaddress]{Millennium Institute of Astrophysics, Chile}
\address[mysecondaryaddress]{Departamento de Ciencias F\'isicas, Universidad Andr\'es Bello, Fern\'andes Concha 700, 759-1598 Las Condes, Santiago, Chile}
\address[mythirdaddress]{Observatorio Astron\'omico de C\'ordoba, Universidad Nacional de C\'ordoba, Laprida 854, X5000BGR, C\'ordoba, Argentina}
\address[myforthaddress]{Instituto de Astrof\'isica, Pontificia Universidad Cat\'olica de Chile, Av. Vicu\~na Mackenna 4860, 782-0436 Macul, Santiago, Chile}
\address[myfifthaddress]{Consejo Nacional de Investigaciones Cient\'ificas y T\'ecnicas (CONICET), Godoy Cruz 2290, CABA, C1425FQB, Argentina}
\address[mysixthaddress]{Unidad de Astronom\'ia, Facultad Cs. B\'asicas, Universidad de Antofagasta, Avda. U. de Antofagasta 02800, Antofagasta, Chile}
\address[myseventhaddress]{Instituto de F\'isica y Astronom\'ia, Facultad de Ciencias, Universidad de Valpara\'iso, Av. Gran Breta\~na 1111, Playa Ancha,
Casilla 5030, Valpara\'iso, Chile}
\address[myeighthaddress]{Universidade Federal do Rio Grande do Sul, Departamento de Astronomia, CP 15051, RS, 91501-970, Porto Alegre, Brazil}

\begin{abstract}
This project is a massive near-infrared (NIR) search for variable stars in highly reddened and obscured open cluster (OC) fields projected on regions of the Galactic bulge and disk. The search is performed using photometric NIR data in the $J$-, $H$- and $K_s$- bands obtained from the Vista Variables in the V\'ia L\'actea (VVV) Survey. We performed in each cluster field a variability search using Stetson's variability statistics to select the variable candidates. Later, those candidates were subjected to a frequency analysis using the Generalized Lomb-Scargle and the Phase Dispersion Minimization algorithms. The number of independent observations range between 63 and 73.  The newly discovered variables in this study, 157 in total in three different known OCs, are classified based on their light curve shapes, periods, amplitudes and their location in the corresponding color-magnitude $(J-K_s,K_s)$ and color-color $(H-K_s,J-H)$ diagrams. We found 5 possible Cepheid stars which, based on the period-luminosity relation, are very likely type II Cepheids located behind the bulge. Among the newly discovered variables, there are eclipsing binaries, $\delta$ Scuti, as well as background RR Lyrae stars. Using the new version of the Wilson \& Devinney code as well as the ``Physics Of Eclipsing Binaries'' (PHOEBE) code, we analyzed some of the best eclipsing binaries we discovered. Our results show that these studied systems turn out to be ranging from detached to double-contact binaries, with low eccentricities and high inclinations of approximately $80^{\circ}$. Their surface temperatures range between $3500$K and $8000$K.
\end{abstract}

\end{frontmatter}


\section{Introduction}

Star clusters are important building blocks of galaxies so knowledge of their individual and statistical properties is of great astrophysical importance. These populations, composed of stars sharing the same age and initial chemical composition, have allowed us to test theories of stellar evolution and have also helped to reveal the structure of their host galaxies. In particular, Galactic open clusters (OCs) have long been considered excellent targets not only to probe the Galactic disk population \citep{friel,bica06} but also to trace its chemical evolution \citep[see, e.g.,][and references therein]{chen}. Estimates indicate that the Milky Way currently hosts a total of about 2.5x10$^4$ or more OCs \citep{port}. However, in the catalogue by \citet[hereafter K13]{k13}, only 2808 Galactic OCs have reasonable estimates of basic cluster parameters such as distance and age, as well as an estimation of the interstellar extinction. This number clearly represents a lower limit to the possible amount of clusters belonging to the Milky Way, if we take into account the recently found clusters  and cluster candidates \citep[see, e.g.,][]{bica03,dutra,borissova11,borissova14,ramirez14,ramirez16,chene,barba}  and the ``still unseen'' OCs, which are deeply embedded in obscured regions or are just too faint to be detected. Distances, masses, and ages for OCs are generally determined from color-magnitude diagrams (CMDs). However, observations of OCs projected on the Galactic bulge region are strongly affected by the effects of both interstellar reddening and high field stellar density \citep{valenti,javier12}. In these objects, the cluster main sequences appear not to be clearly visible, which reduces the accuracy of the relevant physical parameters derived from their CMDs. In these cases, it is very helpful to identify cluster member variable stars since these stars can provide a more precise measurement of the clusters' parameters (particularly their distances). \\

The current project, based on near-infrared (NIR) photometric data, is a search for stellar variability in the fields of Galactic OCs which lie in the highly reddened and obscured regions of the Galactic bulge and disk. As part of a massive search for variable stars in OCs, we characterize for the first time new variable stars detected in three Galactic OC regions located toward the inner Galaxy. For this purpose, we make use of the VISTA Variables in the V\'ia L\'actea (VVV) Survey, which is an ESO Public NIR time-domain survey of the inner Milky Way \citep{minniti}. VVV aims to map the Galactic bulge as well as an adjacent section of the mid-plane, covering stellar populations all the way to the Galactic center, including regions of intense star formation. This survey has been performing time-domain observations in the $K_s$-band for over 5 years \citep{saito12} and provides an atlas of 562 square degrees of the sky in 5 wavebands ($ZYJHK_s$), encompassing about a billion objects. \\

The VVV Survey is discovering hundreds of new clusters, many of them being very distant and deeply embedded objects toward the inner Galaxy \citep[see, e.g.,][]{borissova11,borissova14,chene,ramirez14,ramirez16}. The relevant physical parameters (reddenings, distances, masses, luminosities, sizes and metallicities) for these new clusters are still poorly known or unknown. Our project aims at focusing on all OCs in the VVV survey area. In this first approach, we analyze a few known OCs with previously identified variables. We selected, on the one hand, two moderately young and extended OCs with a relatively high amount of catalogued variables in their fields \citep{zejda} and a considerable amount of new possible variable member candidates. On the other hand, we also selected an intermediate-age more compact OC with only a few known variable stars and some newly discovered variable candidates detected in our study.  Because the clusters are projected onto high density stellar regions of the Galactic bulge, we are at present mainly focused on searching probable NIR counterparts of the already catalogued variable stars as well as searching new variables only discovered using the VVV data. The subsequent goals will be related to the clusters' analysis, i.e.,  to comparing the advantages and disadvantages of analyzing extended and compact OCs and selecting close and faint OCs to warrant their non saturation and usefulness for a future analysis of their proper motions. In addition, with the help of our new data, we aim at improving some of these clusters' parameters that appeared to be uncertain.  \\

In this work we present the first results obtained for three OCs projected on the inner parts of the Galactic bulge, namely Antalova\,1, ASCC\,90 and ESO\,393-15. We describe them in the next section together with the data collected from the VVV Survey. Section 3 details the variable stars found in the fields of these three OCs and their classification based on the obtained light curves. Data analysis and discussion as well as some future work is described in Section 4.  \\

\section{Data acquisition and selected targets}

The observations were made as part of the VVV Survey. The VIRCAM camera on the 4.1\,m VISTA telescope is an array of 16 NIR detectors which produce a combined image of 11.6'$\times$11.6' with a pixel size of 0.34'' \citep{dalton}. The photometry and data reduction have been described in detail elsewhere \citep{saito12,dekany13,javier14}. We briefly mention here that the individual VVV images were reduced, astrometrized and stacked by the Cambridge Astronomy Survey Unit (CASU) using the VISTA Data Flow System (VDFS) pipeline \citep{emerson,irwin,hambly}, and the photometry has been calibrated onto the VISTA filter system. The aperture photometry has been made by CASU on the individual processed images, and generated light curves were then analyzed for variability \citep[see][]{dekany13,javier14}. PSF photometry for each OC in the different available images was extracted and later cross-matched to create the CMDs. A brief description of the selected targets as well as a summary of the previous results for the fields under investigation is given below. \\

\subsection{Antalova\,1}

Antalova\,1 (IAU designation C1725-315) is catalogued as a moderately young and metal-poor OC located in Scorpius at $\alpha_{2000}$ = 17h 28m 57s, $\delta_{2000}$ = -31$^{\circ}$ 34' 48''; l = 355.86$^{\circ}$, b = +1.64$^{\circ}$ ({K13}). It is projected on the inner bulge in the area named b345 in \citet{saito12}. Antalova\,1 has been classified as IV2pn in the Trumpler system \citep{archinal}, i.e., as an OC with the fourth highest concentration degree, a medium range of brightness of its stars and a scanty population. \citet[hereafter K05]{k5a} published a catalogue of astrophysical data for 520 Galactic OCs - among them Antalova\,1 - which could be identified in their All-Sky Compiled Catalogue of 2.5 million stars (ASCC-2.5). By applying homogeneous procedures and algorithms, {K05} determined angular sizes and fundamental astrophysical parameters for their cluster sample. For Antalova\,1, they estimated an angular radius of 35', and obtained the following results: E(B-V) = 0.25, d = 850 pc and 316 Myr. It should be noted, however, that owing to the relatively bright limiting magnitude (V $\sim$ 12.5) of the ASCC-2.5, {K05}'s sample does not include faint and generally remote or highly obscured OCs. More recently, using a combination of uniform kinematic and NIR photometric data gathered in the all-sky catalogue PPMXL \citep{roeser} and the 2MASS catalogue \citep{skrut}, {K13} reported exact positions, apparent radii, proper motions, reddenings, distances and ages for more than 2800 mostly confirmed OCs. For Antalova\,1, they provided the parameters listed in Table \ref{tab1}.  \citet{conrad14} identified Antalova\,1 in the Radial Velocity Experiment \citep[RAVE;][]{stein} and determined its mean cluster metallicity as [M/H] = -0.66 $\pm$ 0.19 (Table \ref{tab1}) from a cleaned working sample. A total of 43 variable stars have been catalogued by \citet[hereafter Z12]{zejda} in the cluster field. \\

\begin{figure*}[ht]
\centering
\includegraphics[width=\hsize]{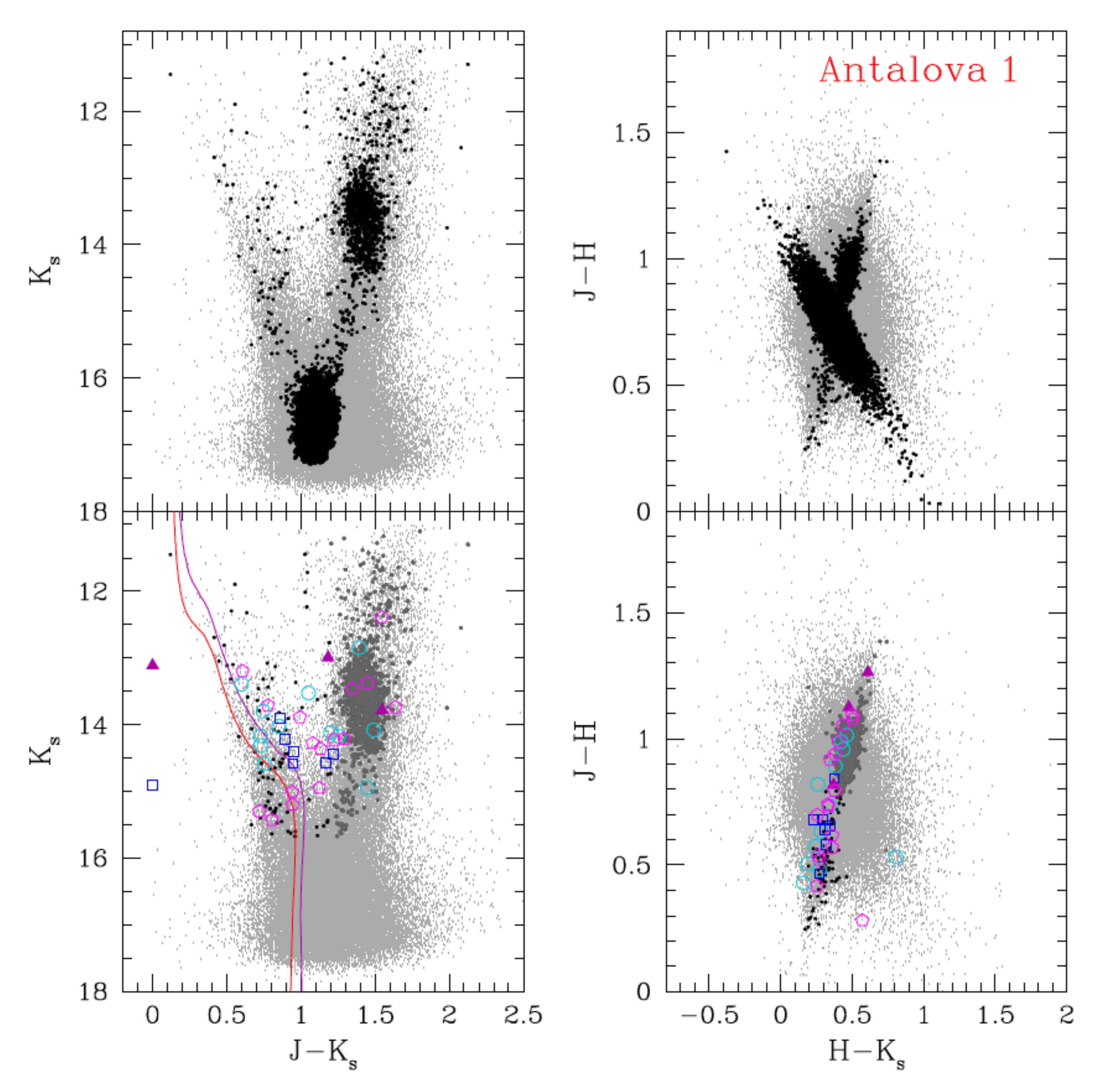}
\caption{Color- magnitude and color-color diagrams of Antalova\,1. Grey dots indicate all stars within the clusters' radii, while black dots represent stars that remain after the decontamination procedure. In the bottom diagrams, we subtracted also the clear bulge main-sequence background and fitted with a red line the \citet{bressan} isochrone corresponding to the parameters estimated by {K13}. The corresponding solar metallicity isochrone is represented by the purple line. Dark grey points represent the bulge red giant branch. We also superimposed on the diagrams the variable stars found in the cluster field. Different symbols and colors represent different types of variables.}
\label{f:cl1}
\end{figure*}

\subsection{ASCC\,90}

ASCC\,90, also known as KPR\,90 \citep{k5b}, has so far received little attention. This cluster is also situated in Scorpius at equatorial coordinates $\alpha_{2000}$ = 17h 39m 07s, $\delta_{2000}$ = -34$^{\circ}$ 48' 54'' and Galactic coordinates l = 354.29$^{\circ}$ and b = -1.91$^{\circ}$. It is projected into the region named b301 in \citet{saito12}. {K05} obtained for ASCC\,90 the following results: $E$($B-V$) = 0.30, d = 500 pc and age = 646 Myr, which are nearly identical to those recently provided by {K13}. {Z12} reported a total of 62 variable stars in the field of ASCC\,90. No metallicity has been determined for this cluster up to now.

\begin{figure*}[ht]
\centering
\includegraphics[width=\hsize]{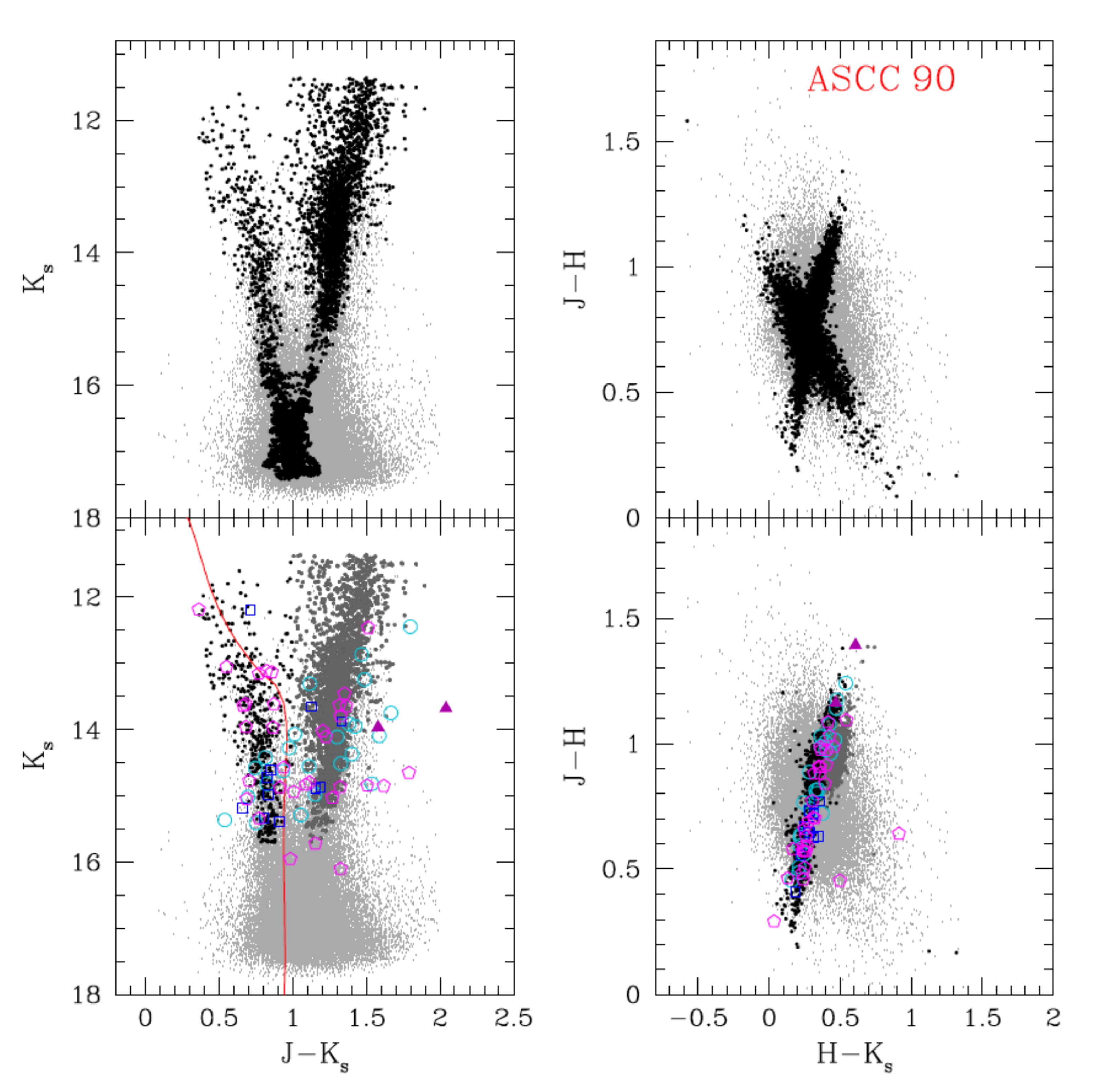}
\caption{Same as Figure \ref{f:cl1} but for ASCC\,90.}
\label{f:cl2}
\end{figure*}

\subsection{ESO\,393-15}

ESO\,393-15 (C1740-342) is an intermediate-age (1.4 Gyr) OC located in a rich stellar field in Scorpius at $\alpha_{2000}$ = 17h 43m 35s, $\delta_{2000}$ = -34$^{\circ}$ 13' 38" and l = 355.26$^{\circ}$ and b = -2.40$^{\circ}$. This is a \citet{trumpler} class III3 OC \citep{archinal}, i.e. a moderately populated cluster with no noticeable concentration and a medium range in the star brightness. ESO\,393-15 lies in the area b302 of \citet{saito12}. According to {K13}, this cluster is a small sized object with an angular diameter of 5.4', located at 2471 pc from the Sun and affected by $E$($B-V$) = 1.67. {Z12} reported five variable stars in the cluster's field. No metallicity estimate has so far been obtained for this cluster. \\

\begin{figure*}[ht]
\centering
\includegraphics[width=\hsize]{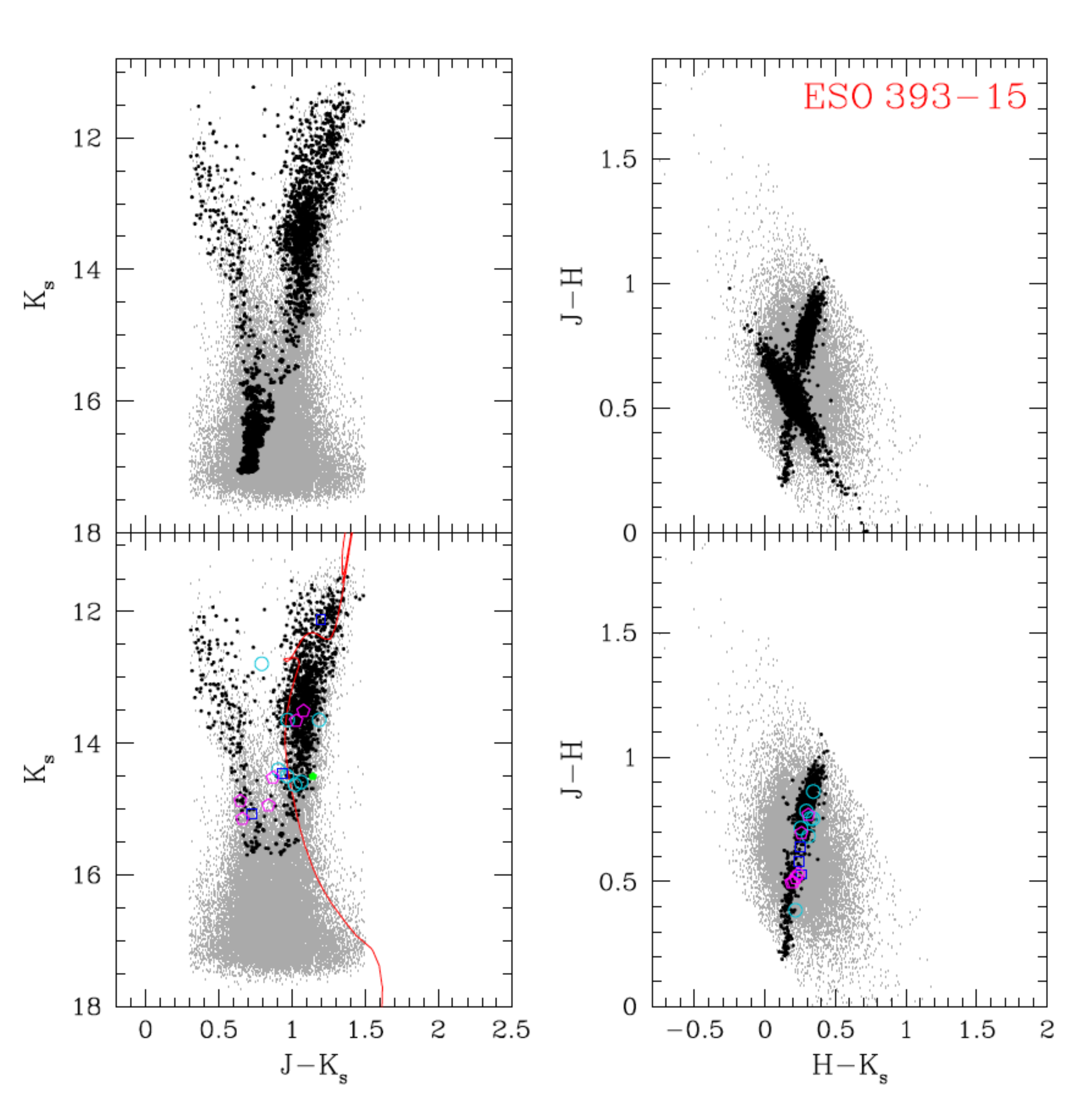}
\caption{Same as Figure \ref{f:cl1} but for ESO\,393-15.}
\label{f:cl3}
\end{figure*}

Table \ref{tab1} lists the coordinates of the clusters and their locations in the VVV area, while Table \ref{tab2} summarizes the fundamental parameters of the selected targets reported by {K13}, including their apparent radii, distances, reddenings, ages, proper motions, radial velocities and metallicities, when available. \\

\begin{table*}
\small
\caption{Open cluster coordinates}
\label{tab1}
\centering
\vspace{0.1cm}
\begin{tabular}{lccccc}
\hline \hline
ID & $\alpha_{2000}$ & $\delta_{2000}$ & l & b  & tile \\
  & [hms] & [dms] & [d] & [d]  & \\
\hline
Antalova\,1 & 17:28:57.0 & -31:34:48 & 355.833 & +01.625  & b345 \\
ASCC\,90    & 17:39:07.2 & -34:48:54 & 354.290 & -01.910  & b301-b302 \\
ESO\,393-15 & 17:43:35.3 & -34:13:38 & 355.261 & -02.396  & b302 \\
\hline
\end{tabular}
\end{table*}

\begin{table}
\small
\caption{Open cluster known parameters taken from \citet{k13}.}
\label{tab2}
\centering
\vspace{0.1cm}
\begin{tabular}{lcccccccc}
\hline \hline
ID & Radius & Dist & $E(J-K_s)$ & $\log t$ & PM$_{\alpha}$ & PM$_{\delta}$ & RV & [Fe/H]  \\
  &  [arcmin] & [pc] & [mag] & [yr] & [mas/yr] & [mas/yr] & [km/s] & [dex]   \\
\hline
Antalova\,1 & 14.1 & 930 & 0.205 & 8.485 & -0.60 & 2.73  & -2.8 & -0.655  \\
ASCC\,90    & 18.3 & 526 & 0.144 & 8.81  & -3.00 & -3.65 & -- & --   \\
ESO\,393-15 & 5.4 & 2471 & 0.800 & 9.15  & -9.15 & 0.23  & -- & --  \\
\hline
\end{tabular}
\end{table}

The $(J-K_s, K_s)$ CMDs as well as $(H-K_s, J-H)$ color-color diagrams were built for each OC. As can be seen in Figures \ref{f:cl1} - \ref{f:cl3}, the clusters are projected onto high stellar density regions where the Galactic disk and bulge populations greatly contaminate the clusters' CMDs (grey dots). A decontamination procedure was required in order to detect and separate clusters from their background stars. With this in mind, we applied a statistical method consisting in defining a ring region surrounding each cluster located 5' away from its field and considered it as a comparison field and counted the stars lying within different intervals of magnitude-color $[\delta K_s, \delta (J-K_s)]$ in the CMD of each selected region. Finally, the number of stars counted for each interval $[\delta K_s, \delta (J-K_s)]$ in the CMD of the comparison fields was subtracted from the number of stars of the corresponding cluster regions. This procedure, applied successfully to other more compact clusters, failed to yield the expected results. It  could be argued that at least in two of the three studied objects, the failure of the cleaning method may be due to the fact that these two OCs are not concentrated and are also very extended, so the separation between cluster and field stars is more confusing. In Figures \ref{f:cl1} - \ref{f:cl3} we marked with black dots the stars  that remain after the decontamination procedure. As can be seen in the upper-right panel, a estrange X-shape appears as resembling two different stellar populations. We believe that the bulge main sequence seen in the color-magnitude diagrams (star concentration at $K_s >$ 16 approximately) is the responsable for such feature. By subtracting this stellar population, the color-magnitude and color-color diagrams of the two lower panels in Figues \ref{f:cl1} - \ref{f:cl3} are obtained. Dark grey points most probably represent the red giant branch of the background bulge older stellar population. Isochrone fittings from \citet{bressan} were superimposed on the CMDs (red lines) based on cluster ages, distances, reddenings, and metallicities (when published) from {K13}. We compare with the reddening maps of \citet{gonzalez11} and \citet{gonzalez12} obtained with the VVV data, but since those maps were calculated for the bulge distances, the values represent an upper limit for the possible clusters reddenings.  \\

\section{Variability analysis}

For each cluster field, we extracted and analyzed VVV data for objects that best matched the positions of previously reported variable stars ({Z12}), and also performed a blind variability search. Objects with putative light variations were selected using Stetson's variability statistics \citep{stetson}. The preselected candidates ($\sim$10\% of the objects) were then subjected to a frequency analysis. Signal detection was performed using the Generalized Lomb-Scargle \citep[GLS;][]{gls} and the Phase Dispersion Minimization \citep[PDM;][]{pdm} algorithms. Phase-folded light curves with the resulting preliminary periods were then visually inspected to select the best solution and to reject spurious signals resulting from various systematics, e.g., rotating diffraction spikes of nearby saturated stars. \\

We refined the periods and optimized the light-curve fit by an iterative procedure including the steps of outlier rejection, aperture optimization, determination of the optimal order of the best-fitting Fourier sum, and refining the period by a non-linear least squares method. The mean apparent $K_s$ magnitudes of the stars and the total amplitudes of the light curves were computed from the final Fourier solutions. At the end of the process, we identified a total of 157 variable stars with the number of independent observations ranging between 63 and 73. \\

\begin{figure}[ht]
\centering
\includegraphics[width=0.7\hsize]{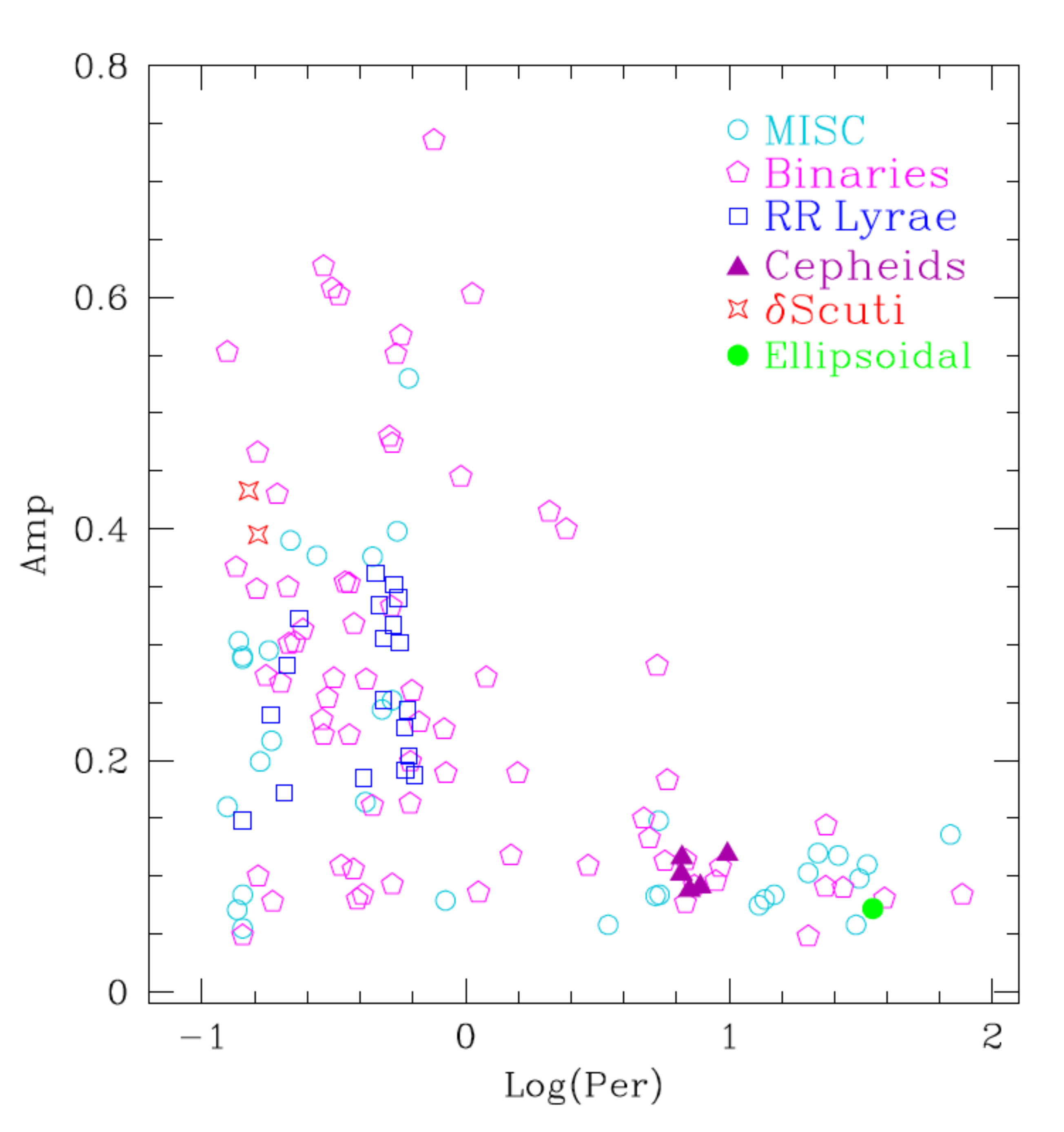}
\caption{Distribution of the new variables in the period-amplitude plane. Different symbols and colors represent the various types of variables.}
\label{f:fig2}
\end{figure}

The classification of periodic variable stars in this study was based on the periods and amplitudes, on visual appearance of the light curves and on the objects' color indices. Classifying NIR light curves is not a straightforward procedure because $K_s$-band light-curves of pulsating variable stars have typically smaller amplitudes and are much more featureless than in the optical region, making it difficult to distinguish different types of objects on the basis of NIR data alone.  \\

The catalogue of {Z12} reports variable stars found in the fields of the three selected clusters. However, by cross-matching their coordinates with the VVV source catalog, we have not found any variable star with NIR counterpart in any of the three fields. We attribute that to the fact that most of the catalogued variables are bright stars, so they are saturated on the VVV $K_s$-band. Other variable stars might also have amplitudes too low to be detected in our procedure as possible variables.  \\

In the field of Antalova\,1, we detected 58 new variable stars from which 7 are newly possible RR Lyrae type stars, 3 were classified as possible Cepheids and 18 as probable eclipsing binary (EB) systems. The remaining ones are reliably classified as undefined types of variable stars (denoted as  MISC or marked with asteriks) or stars whose periods could not be well determined. Their coordinates, periods, amplitudes, $<K_s>$  weighted mean magnitudes from the Fourier fit, $J-K_s$ and $H-K_s$ colors computed from the first $K_s$-band epoch, and classifications are given in Table \ref{t:Anta_var}. In the field of ASCC\,90, we have detected 81 new variable stars. Out of these, 10 are found to be possible RR Lyrae, 2 suspected Cepheids and 1 possible $\delta$ Scuti, while 43 are found to have similar characteristics to EB systems. Their basic properties are detailed in Table \ref{t:ascc_var}. In the field of ESO\,393-15, we identified 18 new variable stars from which 2 are classified as RR\,Lyrae stars, 1 as possible $\delta$ Scuti star, 1 as ellipsoidal and 8 as binary systems. We show in Table \ref{t:eso_var} the corresponding parameters and classifications for all the variable stars found in this field. \\

\begin{figure*}[!ht]
\centering
\includegraphics[width=1.2\hsize]{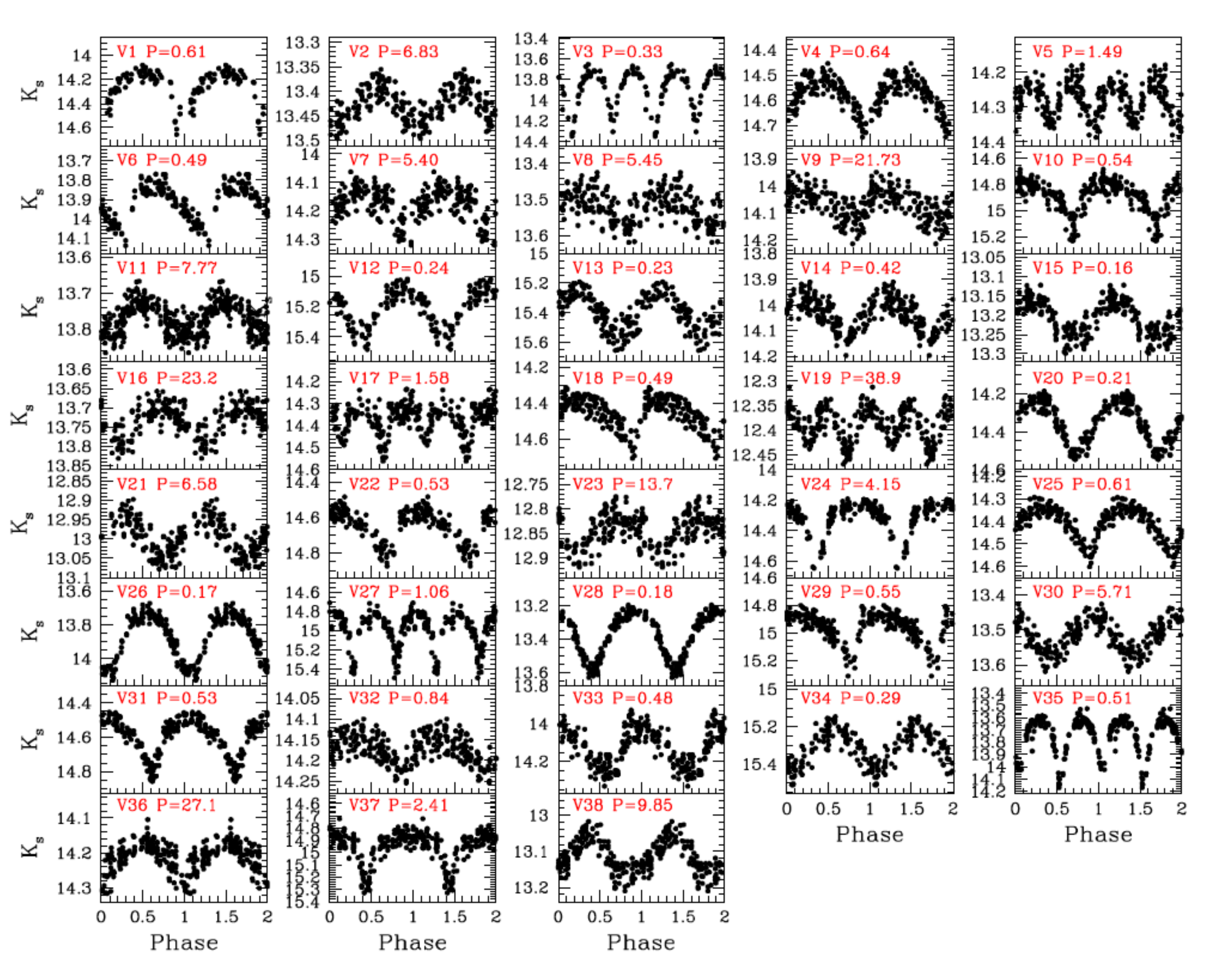}
\caption{Light curves of the variable stars found in the field of Antalova\,1}
\label{f:vlc_cl1}
\end{figure*}

Figure \ref{f:fig2} shows a schematic view in the period-amplitude plane (Bailey diagram) of the whole sample of newly identified variables in the CMDs of the  three  studied OCs. Symbols are the same as in Fig. \ref{f:fig2}. Figures \ref{f:vlc_cl1}-\ref{f:vlc_cl3} show the light curves obtained for all the well classified variables, i.e., variables with Alias periods are not included. \\

\begin{figure*}[!ht]
\centering
\includegraphics[width=1.2\hsize]{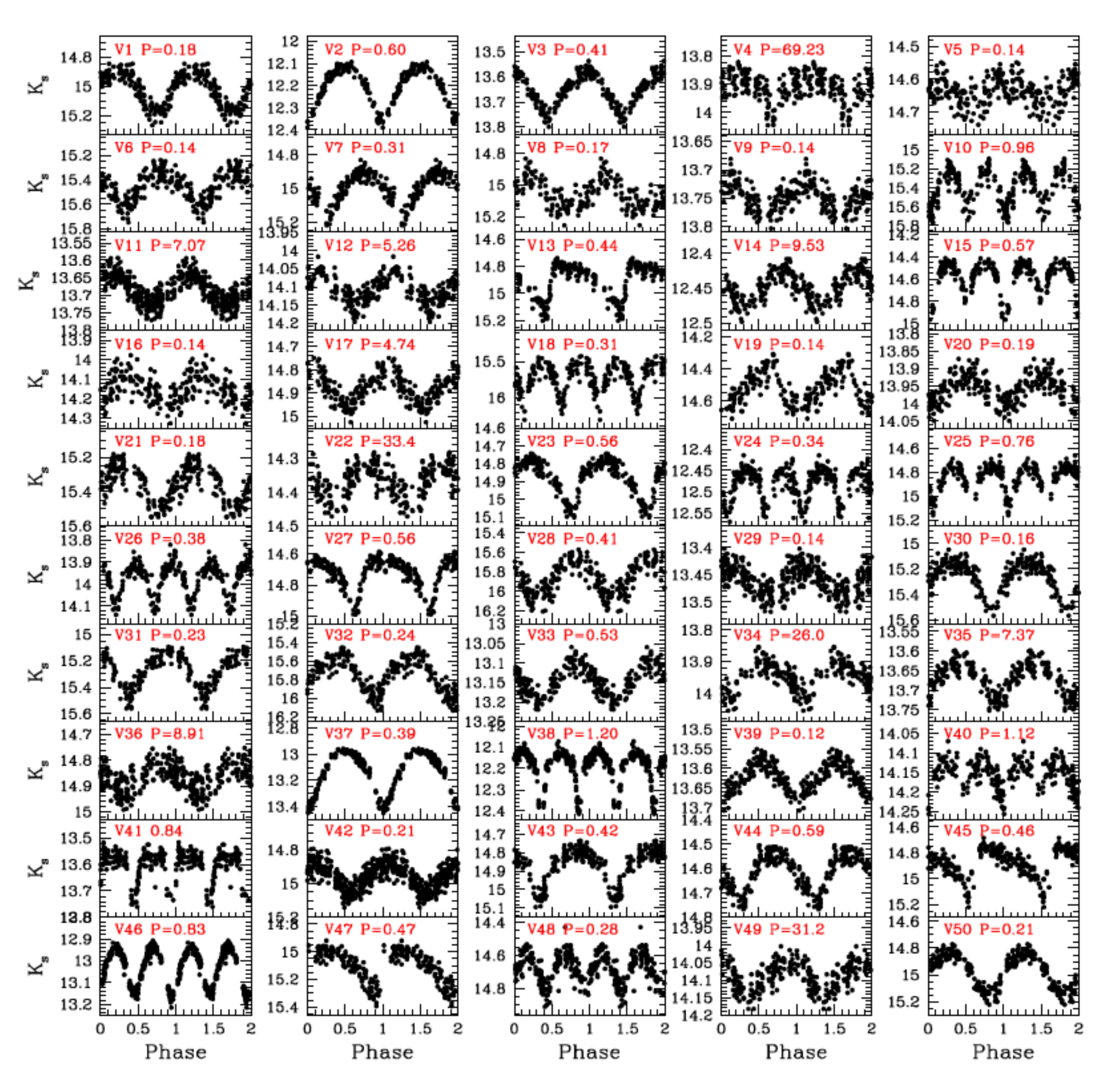}
\caption{Light curves of the variable stars found in the field of ASCC\,90}
\label{f:vlc_cl2_1}
\end{figure*}

\begin{figure*}[!ht]
\setcounter{figure}{5}
\centering
\includegraphics[width=1.1\hsize]{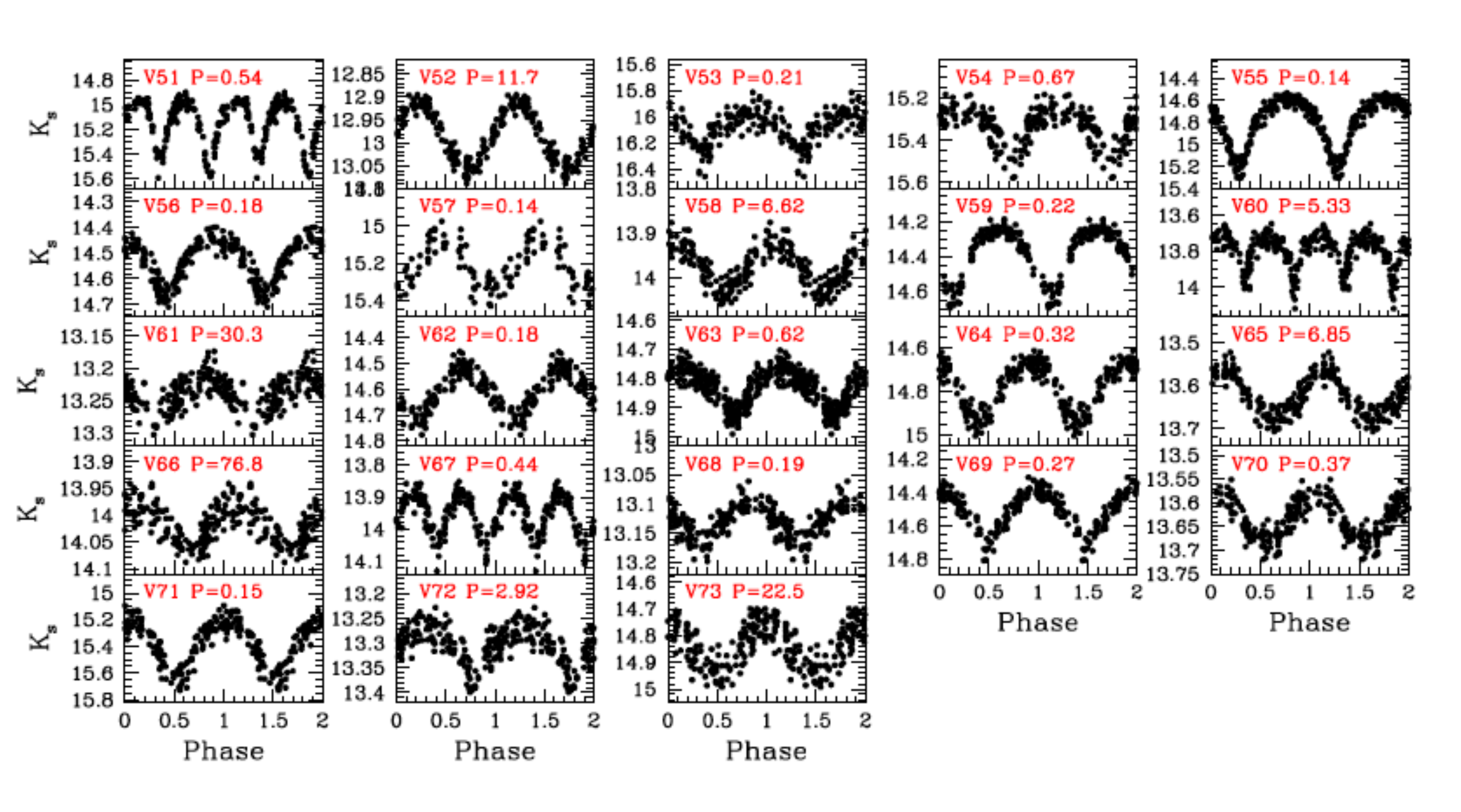}
\caption{Continued.}
\label{f:vlc_cl2_2}
\end{figure*}

\begin{figure*}[!ht]
\centering
\includegraphics[width=1.1\hsize]{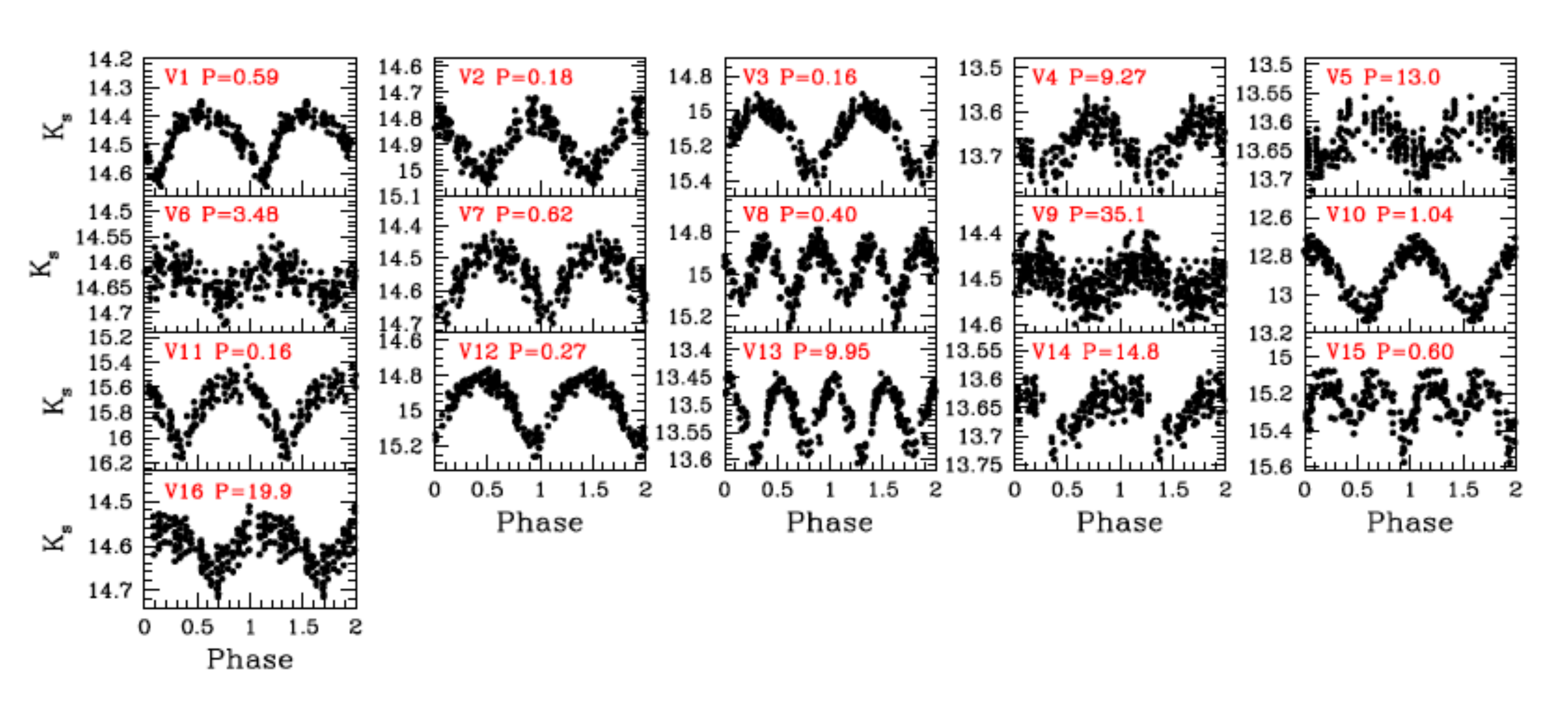}
\caption{Light curves of the variable stars found in the field of ESO\,393-15}
\label{f:vlc_cl3}
\end{figure*}

\newpage
\onecolumn
\begin{flushleft}                                                                   
{\scriptsize
\begin{longtable}{lcccccccccc}
\caption{\label{t:Anta_var} List of variable stars found in Antalova\,1} \\
\hline \hline
 ID &  $\alpha_{2000}$ &  $\delta_{2000}$ &  Dist $^{a}$ &  Period &  Amp $^{b}$ &  $<K_s>$  &  $J-K_s$ &  $H-K_s$ &  Class  &  Comments\\ 
   &  [hms] &  [dms] &  [arcmin] &  [days] &  [mag] &  [mag] &  [mag] &  [mag] &  &  (other types/periods) \\
\noalign{\smallskip}                                                                                                                                                   
\hline
\endfirsthead
\caption{Continued.}\\
\hline \hline
 ID &  $\alpha_{2000}$ &  $\delta_{2000}$ &  Dist $^{a}$ &  Period &  Amp $^{b}$ & $<K_s>$  &  $J-K_s$ &  $H-K_s$ &  Class  &  Comments\\ 
   &  [hms] &  [dms] &  [arcmin] &  [days] &  [mag] &  [mag] &  [mag] &  [mag] &  &  (other types/periods) \\

\noalign{\smallskip}                                                                                                                                                   
\hline
\endhead
\hline

 V1 &  17:28:26.70 &  -31:29:39.0 &  5.17 &  0.60804 &  0.530 &  14.24 &  0.82 &  0.24 &  MISC & . . . \\
  V2 &  17:28:43.23 &  -31:30:50.9 &  10.97 &  6.82885 &  0.077 &  13.43 &  1.49 &  0.44 &  Bin&  Ellipsoidal \\
 V3 &  17:28:46.35 & -31:25:00.9 & 9.79 & 0.33013 & 0.602 & 13.88 & 0.96 & 0.26 & Bin & . . . \\
 V4 &  17:28:47.57 & -31:28:55.0 & 5.89 & 0.63813 & 0.187 & 14.58 & 1.22 & 0.38 & RR* & Bin\\
V5 & 17:28:50.93 & -31:28:55.2 & 5.88 & 1.48478 & 0.118 & 14.28 & 1.07 & 0.33 & Bin & RR; P=0.74298\,d\\
V6 & 17:28:51.09 & -31:28:37.2 & 6.18 & 0.48991 & 0.305 & 13.92 & 0.90 & 0.32 & RR & . . . \\
V7 & 17:28:53.84 & -31:33:35.8 & 1.20 & 5.40317 & 0.148 & 14.18 & 0.81 & 0.27 & MISC & . . . \\
V8 & 17:28:53.87 & -31:31:45.9 & 3.04 & 5.44742 & 0.084 & 13.52 & 1.08 & 0.26 & MISC & Spheroidal* \\
V9 & 17:28:54.68 & -31:23:55.8 & 10.87 & 21.73366 & 0.120 & 14.06 & 1.48 & 0.46 & MISC & . . . \\	
V10 & 17:28:58.37 & -31:22:58.6 & 11.82 & 0.53609 & 0.352 & 14.89 & 0.91 & 0.23 & RR & Bin \\
V11 & 17:29:00.89 & -31:21:39.0 & 13.15 & 7.77096 & 0.091 & 13.77 & 1.60 & 0.48 & Cep* & MISC \\
V12 & 17:29:01.74 & -31:26:39.5 & 8.14 & 0.24126 & 0.313 & 15.21 & 0.85 & 0.57 & Bin & . . . \\
V13 & 17:29:06.70 & -31:31:22.3 & 3.43 & 0.22554 & 0.302 & 15.39 & 0.93 & 0.36 & Bin & . . . \\
V14 & 17:29:11.64 & -31:33:26.8 & 1.38 & 0.41652 & 0.164 & 14.04 & 0.75 & 0.27 & MISC & . . . \\
V15 & 17:29:12.51 & -31:23:24.4 & 11.40 & 0.16319 & 0.100 & 13.21 & 0.67 & 0.25 & Bin & . . . \\	
V16 & 17:29:13.03 & -31:30:21.8 & 4.44 & 23.20854 & 0.091 & 13.73 & 1.60 & 0.50 & Bin & . . . \\
V17 & 17:29:14.26 & -31:27:32.2 & 7.27 & 1.57594 & 0.189 & 14.37 & 0.98 & 0.36 & Bin & . . . \\
V18 & 17:29:14.47 & -31:35:31.0 & 0.77 & 0.48816 & 0.252 & 14.47 & 1.00 & 0.35 & RR & . . . \\
V19 & 17:29:16.09 & -31:27:11.2 & 7.62 & 38.92414 & 0.081 & 12.39 & 1.58 & 0.51 & Bin & . . . \\
V20 & 17:29:16.24 & -31:17:19.1 & 17.48 & 0.21118 & 0.282 & 14.35 & 0.74 & 0.27 & RR & . . . \\	
V21 & 17:29:16.38 & -31:26:55.8 & 7.88 & 6.58358 & 0.102 & 12.99 & 1.19 & 0.37 & Cep* & . . . \\	
V22 & 17:29:16.87 & -31:16:55.7 & 17.87 & 0.52663 & 0.252 & 14.65 & 0.92 & 0.28 & MISC &  . . . \\
V23 & 17:29:17.45 & -31:32:09.3 & 2.67 & 13.66544 & 0.080 & 12.85 & 1.40 & 0.44 & MISC & . . . \\
V24 & 17:29:18.23 & -31:25:38.5 & 9.17 & 2.07537 & 0.415 & 14.33 & 1.17 & 0.38 & Bin & . . . \\	
V25 & 17:29:18.29 & -31:35:42.0 & 0.97 & 0.60952 & 0.204 & RR & 14.41 & 0.97 & 0.29 & Bin \\
V26 & 17:29:18.75 & -31:21:26.7 & 13.36 & 0.17437 & 0.354 & 13.87 & 0.69 & 0.19 & MISC & Bin; P=0.34874\\
V27 & 17:29:20.22 & -31:35:36.5 & 0.90 & 1.06129 & 0.603 & 14.96 & 1.07 & 0.33 & Bin & P=0.53066 \\
V28 & 17:29:21.23 & -31:21:43.8 & 13.08 & 0.18109 & 0.353 & 13.36 & 0.59 & 0.16 & MISC & Bin; P=0.36216 \\
V29 & 17:29:22.00 & -31:31:07.1 & 3.71 & 0.55032 & 0.398 & 14.97 & 1.27 & 0.38 & MISC & . . . \\
V30 & 17:29:22.17 & -31:35:01.1 & 0.47 & 5.70666 & 0.113 & 13.52 & 1.32 & 0.38 & Bin & . . . \\
V31 & 17:29:24.16 & -31:28:27.7 & 6.35 & 0.53211 & 0.317 & 14.59 & 0.95 & 0.31 & RR/Bin & . . . \\	
V32 & 17:29:26.25 & -31:34:25.8 & 0.61 & 0.83882 & 0.079 & 14.17 & 1.41 & 0.42 & MISC  & . . . \\
V33 & 17:29:29.30 & -31:37:16.6 & 2.53 & 0.48178 & 0.244 & 14.12 & 1.08 & 0.34 & MISC & . . . \\
V34 & 17:29:29.90 & -31:30:52.7 & 3.96 & 0.28913 & 0.222 & 15.31 & 0.79 & 0.27 & Bin & MISC \\
V35 & 17:29:31.07 & -31:25:01.1 & 9.80 & 0.51328 & 0.480 & 13.82 & 0.80 & 0.26 & Bin & . . . \\	
V36 & 17:29:31.57 & -31:36:04.7 & 1.40 & 27.07240 & 0.090 & 14.21 & 1.40 & 0.40 & Bin & MISC\\
V37 & 17:29:34.09 & -31:33:05.4 & 1.82 & 2.40947 & 0.400 & 14.95 & 1.26 & 0.35 & Bin & . . . \\
V38 & 17:29:35.26 & -31:34:58.1 & 0.66 & 9.85270 & 0.119 & 13.11 & 1.87 & 0.61 & Cep* & . . . \\
\hline 
V39 & 17:28:29.67 & -31:29:05.4 & 5.73 & . . . & 0.236 & 13.93 &  1.86 & 0.61 & MISC & . . . \\
V40 & 17:28:32.97 & -31:30:24.2 & 4.41 & . . . & 0.305 & 14.43 & 1.80 & 0.63 & MISC & . . . \\
V41 & 17:28:53.86 & -31:28:31.8 & 6.27 & . . .	& 0.227 & 14.90 & 1.40 & 0.43 & MISC & . . . \\
V42 & 17:28:58.92 & -31:21:40.4 & 13.13 & . . . & 0.162 & 12.74 & 0.69 & 0.34 & MISC & . . .  \\	
V43 & 17:29:01.23 & -31:34:01.9 & 0.77 & . . .	& 0.306 & 14.19 & 1.23 & 0.44 & MISC & . . . \\
V44 & 17:29:02.51 & -31:33:52.3 & 0.93 & . . . & 0.298 & 12.99 & 1.61 & 0.57 & MISC & . . . \\
V45 & 17:29:03.44 & -31:20:31.7 & 14.27 & . . . & 0.110 & 12.99 & 1.46 & 0.50 & MISC & . . . \\	
V46 & 17:29:03.67 & -31:33:37.5 & 1.18 & . . .	& 0.387 & 14.72 & 1.25 & 0.50 & MISC & . . . \\
V47 & 17:29:05.37 & -31:30:22.0 & 4.44 & . . . & 0.303 & 14.34 & 1.73 & 0.59 & MISC & . . . \\
V48 & 17:29:06.48 & -31:31:57.1 & 2.85 & . . .	& 0.278 & 14.28 & 1.08 & 0.31 & MISC & . . . \\
V49 & 17:29:09.50 & -31:28:01.6 & 6.78 & . . .	& 0.346 & 14.67 & 1.43 & 0.59 & MISC & . . . \\
V50 & 17:29:11.58 & -31:34:18.9 & 0.54 & . . .	& 0.561 & 13.77 & 2.06 & 0.71 & MISC & . . . \\
V51 & 17:29:13.84 & -31:21:55.6 & 12.88 & . . . & 0.257 & 13.76 & 1.36 & 0.39 & MISC & . . . \\
V52 & 17:29:2.84 & -31:34:28.6 & 0.57 & . . .	& 0.132 & 14.44 & 1.27 & 0.45 & MISC &  . . . \\
V53 & 17:29:25.01 & -31:31:54.8 & 2.92 & . . . & 0.230 & 14.45 & 1.29 & 0.41 & MISC & . . . \\
V54 & 17:29:25.11 & -31:34:52.0 & 0.47 & . . .	& 0.443 & 15.01 & 1.65 & . . . & MISC & . . . \\
V55 & 17:29:29.68 & -31:21:45.2 & 13.06 & . . . & 0.205 & 14.48 & . . . & . . . & MISC & . . . \\	
V56 & 17:29:31.32 & -31:22:33.1 & 12.26 & . . . & 0.219 & 13.95 & 0.85 & 0.41 & MISC & . . . \\
V57 & 17:29:34.02 & -31:38:55.2 & 4.17 & . . . & 0.130 & 14.14 & 1.08 & 0.36 & MISC & . . .\\
V58 & 17:29:35.60 & -31:31:19.2 & 3.54 & . . . & 0.430 & 14.24 & 0.96 & 0.27 & MISC & . . . \\	
\hline
\end{longtable}
\medskip
$^{a}$ Distance to the cluster center. $^{b} K_s$ total amplitude. }\\ 

\onecolumn
{\scriptsize
\begin{longtable}{lcccccccccc}
\caption{\label{t:ascc_var} List of variable stars found in ASCC\,90} \\
\hline \hline
ID & $\alpha_{2000}$ & $\delta_{2000}$ & Dist $^{a}$ & Period & Amp $^{b}$ & <$K_s$>  & $J-K_s$ & $H-K_s$ & Class  & Comments\\ 
   & [hms] & [dms] & [arcmin] & [days] & [mag] & [mag] & [mag] & [mag] &  & (other types/periods) \\
\hline
\endfirsthead
\caption{Continued.}\\
\hline \hline
ID & $\alpha_{2000}$ & $\delta_{2000}$ & Dist $^{a}$ & Period & Amp $^{b}$ & <$K_s$>  & $J-K_s$ & $H-K_s$ & Class  & Comments\\ 
   & [hms] & [dms] & [arcmin] & [days] & [mag] & [mag] & [mag] & [mag] &  & (other types/periods) \\
\hline
\endhead
\hline
V1 & 17:37:56.45 & -34:49:51.1 & 17.71 & 0.17500 & 0.273 & 15.02 & 0.75 & 0.17 & Bin & $\delta$ Sct \\
V2 & 17:37:57.42 & -34:50:46.3 & 17.55 & 0.60235 & 0.244 & 12.21 & 0.59 & 0.18 & RR/Bin & . . . \\
V3 & 17:38:00.24 & -34:43:10.0 & 17.69 & 0.41059 & 0.185 & 13.65 & 1.02 & 0.28 & RR/Bin & . . . \\
V4 & 17:38:04.41 & -34:49:15.8 & 15.70 & 69.23252 & 0.136 & 13.91 & 1.39 & 0.43 & MISC & . . . \\
V5 & 17:38:06.23 & -34:44:11.2 & 15.96 & 0.13602 & 0.071 & 14.64 & 0.81 & 0.22 &  MISC & . . . \\
V6 & 17:38:09.52 & -34:41:55.3 & 16.02 & 0.13791 & 0.303 & 15.45 & 1.17 & 0.28 & MISC & . . . \\
V7 & 17:38:12.52 & -34:50:02.1 & 13.72 & 0.31308 & 0.260 & 15.00 & 0.69 & 0.21 & MISC & Bin; P=0.62616 \\
V8 & 17:38:16.80 & -34:41:22.3 & 14.68 & 0.16621 & 0.199 & 15.05 & 1.40 & 0.40 & MISC & . . . \\
V9 & 17:38:16.81 & -34:40:58.5 & 7.07 & 0.14247 & 0.055 & 13.75 & 1.65 & 0.48 & MISC & . . . \\
V10 & 17:38:23.72 & -34:48:19.2 & 10.89 & 0.47946 & 0.445 & 15.43 & 0.72 & 0.21 & MISC & Bin; P=0.95893 \\
V11 & 17:38:27.14 & -34:35:38.8 & 16.62 & 7.06969 & 0.088 & 13.68 & 2.00 & 0.61 & Cep* & . . . \\
V12 & 17:38:28.74 & -34:40:22.0 & 12.86 & 5.26004 & 0.083 & 14.10 & 1.35 & 0.36 & MISC & . . . \\
V13 & 17:38:30.78 & -34:44:38.8 & 10.05 & 0.44373 & 0.376 & 14.90 & 1.00 & 0.24 & MISC & . . . \\
V14 & 17:38:32.66 & -34:35:32.7 & 15.91 & 9.52668 & 0.048 & 12.45 & 1.78 & 0.54 & MISC & Bin; P=19.04586 \\
V15 & 17:38:34.82 & -34:33:26.9 & 17.44 & 0.56796 & 0.567 & 14.63 & 1.02 & 0.32 & Bin & . . . \\
V16 & 17:38:35.83 & -34:38:45.9 & 12.81 & 0.14247 & 0.148 & 14.15 & 1.04 & 0.32 & RR & $\delta$ Sct* \\
V17 & 17:38:41.52 & -34:45:55.7 & 7.07 & 4.73548 & 0.150 & 14.88 & 1.35 & 0.36 & Bin & . . . \\
V18 & 17:38:41.67 & -34:33:26.0 & 16.73 & 0.31084 & 0.608 & 15.72 & 1.20 & 0.32 & Bin & . . . \\
V19 & 17:38:42.87 & -34:49:42.6 & 16.73 & 0.14247 & 0.290 & 14.52 & 1.14 & 0.33 & MISC & . . . \\
V20 & 17:38:43.33 & -34:36:14.5 & 14.00 & 0.19380 & 0.080 & 13.96 & 1.03 & 0.29 & MISC & Bin; P=0.38759 \\
V21 & 17:38:43.55 & -34:46:41.7 & 6.31 & 0.17915 & 0.295 & 15.36 & 0.85 & 0.22 & MISC & . . . \\
V22 & 17:38:46.21 & -34:32:11.7 & 17.51 & 33.44160 & 0.110 & 14.37 & 1.39 & 0.36 & MISC & . . . \\
V23 & 17:38:49.38 & -34:44:55.2 & 5.97 & 0.56232 & 0.302 & 14.88 & 1.13 & 0.36 & RR* & . . . \\
V24 & 17:38:49.64 & -34:44:42.2 & 6.07 & 0.33683 & 0.109 & 12.45 & 1.46 & 0.42 &  Bin &  P=0.16841 \\
V25 & 17:38:51.49 & -34:43:57.4 & 6.31 & 0.75759 & 0.648 & 14.87 & 0.30 & 1.08 &  Bin & . . . \\
V26 & 17:38:52.06 & -34:46:41.9 & 11.29 & 0.37722 & 0.220 & 14.00 & 0.95 & 0.50 & Bin &  P=0.18861  \\
V27 & 17:38:52.21 & -34:34:44.1 & 14.65 & 0.55657 & 0.340 & 14.73 & 0.95 & 0.28 & RR* & Bin \\
V28 & 17:38:54.31 & -34:49:44.8 & 3.33 & 0.40672 & 0.390 & 15.85 & 0.68 & 0.17 & Bin & . . . \\
V29 & 17:38:57.99 & -34:49:25.8 & 2.36 & 0.14247 & 0.049 & 13.46 & 1.34 & 0.36 & Bin* & . . . \\
V30 & 17:38:59.59 & -34:34:20.5 & 14.68 & 0.16144 & 0.348 & 15.27 & . . . & 0.26 & Bin & . . . \\
V31 & 17:39:01.16 & -34:33:12.2 & 15.77 & 0.23383 & 0.322 & 15.26 & 0.78 & 0.22 & RR* & . . . \\
V32 & 17:39:01.84 & -34:49:41.6 & 15.68 & 0.23723 & 0.474 & 15.73 & 0.84 & 0.24 & MISC & . . . \\
V33 & 17:39:03.03 & -34:34:15.3 & 14.68 & 0.52639 & 0.093 & 13.14 & 0.81 & 0.24 & Bin & . . .\\
V34 & 17:39:06.21 & -34:41:41.7 & 7.21 & 25.96282 & 0.118 & 13.95 & 1.61 & 0.47 & MISC & . . . \\
V35 & 17:39:10.17 & -34:31:43.1 & 17.20 & 7.36927 & 0.092 & 13.67 & 1.38 & 0.40 & Bin & . . .\\
V36 & 17:39:12.84 & -34:44:37.6 & 4.50 & 8.91203 & 0.096 & 14.86 & 1.50 & 0.41 & Bin & . . . \\
V37 & 17:39:13.47 & -34:40:25.7 & 8.61 & 0.19258 & 0.430 & 13.13 & 0.89 & 0.26 & Bin & . . . \\
V38 & 17:39:14.85 & -34:41:43.9 & 7.42 & 1.19758 & 0.272 & 12.19 & 0.33 & 0.03 & Bin & P=0.59881 \\
V39 & 17:39:15.71 & -34:31:38.1 & 17.40 & 0.12466 & 0.553 & 13.62 & 0.91 & 0.27 & Bin* & . . . \\
V40 & 17:39:21.22 & -34:40:02.3 & 9.53 & 1.11931 & 0.086 & 14.10 & 1.23 & 0.39 & Bin & MISC; P=0.55964 \\
V41 & 17:39:21.62 & -34:44:08.5 & 5.97 & 0.84096 & 0.189 & 13.63 & 0.70 & 0.24 & Bin & . . . \\
V42 & 17:39:22.93 & -34:43:02.0 & 7.06 & 0.20557 & 0.172 & 14.95 & 0.92 & 0.25 & RR* & . . . \\ 
V43 & 17:39:24.41 & -34:36:38.5 & 12.99 & 0.41902 & 0.270 & 14.87 & 1.02 & 0.31 & Bin & . . .\\
V44 & 17:39:29.49 & -34:47:04.7 & 5.86 & 0.59093 & 0.192 & 14.62 & 0.93 & 0.29 & RR & Bin\\
V45 & 17:39:31.18 & -34:40:08.5 & 10.61 & 0.45602 & 0.362 & 14.88 & 1.04 & 0.32 & RR* & . . . \\
V46 & 17:39:31.58 & -34:43:44.3 & 7.99 & 0.82798 & 0.227 & 13.03 & 0.59 & 0.13 & Bin & P=0.41489\,d \\
V47 & 17:39:33.43 & -34:48:00.6 & 6.62 & 0.46938 & 0.334 & 15.08 & 0.97 & 0.34 & RR & . . . \\
V48 & 17:39:34.52 & -34:37:47.0 & 13.05 & 0.28486 & 0.235 & 14.66 & 1.64 & 0.54 & Bin & . . . \\
V49 & 17:39:39.15 & -34:39:26.3 & 12.38 & 31.21711 & 0.098 & 14.09 & 1.50 & 0.42 & MISC & . . . \\ 
V50 & 17:39:39.48 & -34:35:06.6 & 15.98 & 0.21359 & 0.301 & 14.97 & 0.86 & 0.27 & Bin & . . . \\
V51 & 17:39:43.53 & -34:43:29.1 & 10.57 & 0.54353 & 0.551 & 15.16 & 0.83 & 0.26 & Bin & RR \\
V52 & 17:39:47.16 & -34:40:15.5 & 13.21 & 11.67808 & 0.144 & 12.98 & 1.37 & 0.37 & MISC & Bin; P=23.34019 \\
V53 & 17:39:47.51 & -34:43:31.5 & 11.42 & 0.21146 & 0.350 & 16.07 & 1.24 & 0.36 & Bin & . . . \\ 
V54 & 17:39:47.64 & -34:48:30.7 & 10.12 & 0.66579 & 0.233 & 15.34 & 0.81 & 0.24 & Bin & . . . \\
V55 & 17:39:48.19 & -34:46:30.2 & 10.53 & 0.14445 & 0.627 & 14.77 & 0.86 & 0.22 & MISC & Bin;  P=0.28891\\
V56 & 17:39:48.47 & -34:38:39.6 & 14.53 & 0.18105 & 0.233 & 14.52 & 14.52 & 0.34 & MISC & Bin; P=0.36210 \\
V57 & 17:39:48.60 & -34:39:40.6 & 13.86 & 0.14247 & 0.288 & 15.20 & . . . & . . . &  MISC & . . . \\
V58 & 17:39:49.25 & -34:38:31.3 & 14.77 & 6.62427 & 0.116 & 13.97 & 1.63 & 0.47 & Cep* & . . .  \\
V59 & 17:39:51.82 & -34:49:31.2 & 11.17 & 0.21663 & 0.390 & 14.38 & 0.80 & 0.24 & MISC & . . . \\
V60 & 17:39:57.00 & -34:45:38.3 & 8.61 & 5.32970 & 0.282 & 13.94 & 1.27 & 0.36 & Bin & . . . \\
V61 & 17:39:57.11 & -34:44:32.7 & 13.21 & 30.33833 & 0.058 & 13.24 & 3.29 & 2.50 & MISC & . . . \\
V62 & 17:39:58.25 & -34:38:45.5 & 16.30 & 0.18358 & 0.217 & 14.61 & 1.10 & 0.37 & MISC & . . . \\
V63 & 17:39:59.22 & -34:47:07.6 & 13.22 & 0.61524 & 0.163 & 14.83 & 0.97 & 0.28 & Bin & . . . \\
V64 & 17:40:01.28 & -34:40:45.6 & 15.78 & 0.31585 & 0.271 & 14.78 & 0.82 & 0.25 & Bin & . . . \\
V65 & 17:40:01.57 & -34:36:49.8 & 18.18 & 6.85206 & 0.114 & 13.62 & 1.32 & 0.40 & Bin & . . . \\
V66 & 17:40:04.50 & -34:45:33.9 & 14.70 & 76.75488 & 0.084 & 14.02 & 1.26 & 0.35 & Bin & . . . \\
V67 & 17:40:04.78 & -34:39:55.4 & 16.96 & 0.44270 & 0.161 & 13.96 & 0.76 & 0.25 & Bin & P=0.22135 \\
V68 & 17:40:04.89 & -34:45:57.9 & 14.72 & 0.18533 & 0.078 & 13.13 & 0.91 & 0.25 & Bin & . . . \\
V69 & 17:40:05.48 & -34:45:31.4 & 14.96 & 0.27276 & 0.377 & 14.51 & 0.61 & 0.16 & MISC & . . . \\
V70 & 17:40:07.25 & -34:41:46.2 & 16.62 & 0.37462 & 0.106 & 13.63 & 0.72 & 0.24 & Bin & . . . \\
V71 & 17:40:07.59 & -34:40:28.1 & 17.21 & 0.15041 & 0.433 & 15.38 & 0.93 & 0.31 & $\delta$ Sct & . . . \\
V72 & 17:40:09.37 & -34:43:47.9 & 16.36 & 2.91559 & 0.109 & 13.30 &1.14 & 0.33 & MISC & Bin; P=5.83089 \\
V73 & 17:40:13.58 & -34:43:44.7 & 17.38 & 5.83096 & 0.183 & 14.84 & 1.44 & 0.46 & Bin & P=22.45679\\
\hline
V74 & 17:38:22.24 & -34:21:45.7 & 11.01 & . . . & 0.170 & 14.62 & 1.11 & 0.28 & MISC & . . . \\
V75 & 17:38:22.95 & -34:46:38.9 & 11.29 & . . . & 0.066 & 13.51 & 1.19 & 0.31 & MISC & . . . \\
V76 & 17:38:45.39 & -34:27:47.4 & 18.22 & . . . & 0.594 & 15.47 & 1.96 & 0.58 & MISC & . . . \\
V77 & 17:38:55.04 & -34:30:07.7 & 14.04 & . . . & 0.315 & 15.01 & 1.39 & 0.36 & MISC & . . . \\
V78 & 17:38:57.45 & -34:29:18.0 & 14.72 & . . . & 0.238 & 15.16 & . . . & 0.56 & MISC & . . . \\
V79 & 17:39:22.98 & -34:29:16.7 & 16.41 & . . . & 0.250 & 14.19 & 1.29 & . . . & MISC & . . . \\
V80 & 17:39:26.48 & -34:40:26.5 & 7.84 & . . . & 0.464 & 15.03 & 1.58 & 0.41 &  MISC & . . . \\
V81 & 17:39:28.21 & -34 43:05.3 & 14.71 & . . . & 0.080 & 13.52 & 1.33 & 0.40 & MISC & . . .  \\
\hline
\end{longtable}
\medskip
$^{a}$ Distance to the cluster center. $^{b} K_s$ total amplitude. }\\ 

\onecolumn
{\scriptsize
\begin{longtable}{lcccccccccc}
\caption{\label{t:eso_var} List of variable stars found in ESO\,393-15} \\
\hline \hline
ID & $\alpha_{2000}$ & $\delta_{2000}$ & Dist $^{a}$ & Period & Amp $^{b}$ & <$K_s$>  & $J-K_s$ & $H-K_s$ & Class  & Comments\\ 
   & [hms] & [dms] & [arcmin] & [days] & [mag] & [mag] & [mag] & [mag] &  & (other types/periods) \\
\hline
V1 & 17:43:14.58 & -34:13:19.2 & 5.19 & 0.58904 & 0.228  & 14.47 & 0.79 & 0.26 & RR* & . . . \\
V2 & 17:43:19.36 & -34:10:48.6 & 4.88 & 0.18203 & 0.221 & 14.89 & 0.70 & 0.20 & Bin & . . . \\
V3 & 17:43:21.49 & -34:14:23.1 & 3.53 & 0.16302 & 0.395 & 15.13 & 0.82 & 0.24 & $\delta$ Sct & . . . \\
V4 & 17:43:22.09 & -34:12:10.9 & 3.61 & 9.26665 & 0.108 & 13.67 & 0.95 & 0.26 & Bin* & . . . \\
V5 & 17:43:22.45 & -34:11:42.9 & 3.74 & 12.98331 & 0.075 & 13.63 & 1.20 & 0.34 & MISC & . . . \\
V6 & 17:43:24.76 & -34:13:08.6 & 2.68 & 3.47711 & 0.058 & 14.64 & 1.08 & 0.29 & MISC & . . . \\
V7 & 17:43:25.80 & -34:16:42.2 & 3.88 & 0.61680 & 0.199 & 14.55 & 0.77 & 0.24 & Bin/RR & . . . \\
V8 & 17:43:25.81 & -34:13:31.4 & 2.38 & 0.39748 & 0.267 & 14.97 & 0.73 & 0.22 & Bin & . . . \\
V9 & 17:43:31.30 & -34:10:07.5 & 3.65 & 35.12639 & 0.072 & 14.50 & 1.19 & 0.30 & Ellipsoidal* & . . . \\
V10 & 17:43:33.39 & -34:13:43.8 & 0.49 & 0.52252 & 0.333 & 12.91 & 0.60 & 0.22 & MISC & Bin; P=1.04504  \\
V11 & 17:43:36.01 & -34:15:55.0 & 2.29 & 0.16241 & 0.466 & 15.73 & 1.66 & 0.31 & Bin & . . . \\
V12 & 17:43:36.62 & -34:14:35.0 & 1.01 & 0.13454 & 0.367 & 14.95 & 0.89 & 0.25 & RR & Bin; P=0.26908  \\
V13 & 17:43:36.74 & -34:17:06.4 & 3.49 & 9.95217 & 0.133 & 13.52 & 1.08 & 0.31 & Bin &  P=4.97596 \\
V14 & 17:43:39.01 & -34:12:08.9 & 1.75 & 14.83956 & 0.084 & 13.65 & 0.99 & 0.31 & MISC & . . . \\
V15 & 17:43:40.77 & -34:16:20.5 & 3.03 & 0.59805 & 0.254 & 15.24 & 0.68 & 0.19 & Bin & . . . \\
V16 & 17:43:44.46 & -34:11:08.7 & 3.38 & 19.94113 & 0.103 & 14.60 & 1.07 & 0.32 & MISC & . . . \\
\hline
V17 & 17:43:29.27 & -34:15:50.0 & 2.67 & . . . & 0.261 & 13.89 & 1.19 & 0.39 & MISC & . . . \\
V18 & 17:43:33.04 & -34:09:41.2 & 3.99 & . . . & 0.227 & 14.59 & 1.22 & 0.49 & MISC & . . . \\
\hline
\end{longtable}
\medskip
$^{a}$ Distance to the cluster center. $^{b} K_s$ total amplitude. }\\ 

\end{flushleft}

\section{Discussion}

Antalova\,1 and ASCC\,90 are moderately young OCs, while ESO\,393-15 is an intermediate-age OC (Table \ref{tab1}). Therefore, it is quite clear that the RR Lyrae stars found in their respective fields must be background stars projected in the direction of the clusters. However, since our project aims at searching variable stars as a whole and in the Galactic OC fields in particular, the classification of such new variables as RR Lyrae stars enlarges not only the sample of RR Lyrae variables currently known but also the background stellar properties to be studied from them. \\

\subsection{Cepheids analysis}

We have detected a total of 5 Cepheid candidates in the fields of Antalova\,1 and ASCC\,90. In order to determine their probability of being cluster members, we calculated their distances and interstellar extinctions by using the period-luminosity (PL) relations in the $H$ and $K_s$ passbands. To determine the type of each Cepheid, we used the PL relations for classical and type II Cepheids. Assuming at first they are classical Cepheids, we used the following PL relations of \citet{dekany15} adapted to the VVV Survey passbands:
\begin{equation}
M_H = -3.228\,[\pm0.06] \times (\log P - 1) - 5.617\,[\pm0.048],
\end{equation}
\begin{equation}
M_{K_s} = -3.269\,[\pm0.05] \times (\log P - 1) - 5.663\,[\pm0.048], 
\end{equation}
\noindent where $M_H$ and $M_{K_s}$ are the absolute magnitudes in the $H$ and $K_s$ bands, respectively.  The color excesses are calculated from:
\begin{equation}
E(H-K_s) = < H-K_s> - (M_H-M_{K_s}), 
\end{equation}
\noindent where $<H-K_s>$ is the mean $(H-K_s)$ color index as defined in \citet{dekany15}. The total absorption in the $Ks$-band and the individual distances R for both Cepheid types were derived from the following relations given by \citet{dekany15}:
\begin{equation}
A(K_s) = 1.634 \times E(H-K_s),
\end{equation}
\begin{equation}
\log R = 1 + 0.2 \times (<K_s> - A(K_s)-M_{K_s}),
\end{equation}

\noindent where the mean $<K_s>$ magnitudes of the stars were computed from the optimized Fourier fits of the light curves. Secondly, assuming the stars are type II Cepheids, the following PL relations from \citet{matsunaga} were employed:
\begin{equation}
M_H = -2.340\,[\pm0.05] \times (\log P - 1.2) + 14.760\,[\pm0.017],
\end{equation}
\begin{equation}
M_{K_s} = -2.410\,[\pm0.05] \times (\log P - 1.2) + 14.617\,[\pm0.015]. 
\end{equation}

The results obtained after an analysis of the possible types of Cepheids are shown in Table \ref{t:cep}. Since types I and II Cepheids have similar periods and NIR light curves, it is difficult to distinguish their right types in a first approach. Taking into account the  $K_s$ reddening map of the VVV bulge obtained by \citet{gonzalez11} and \citet{gonzalez12}, a detailed analysis was recently carried out by \citet{dekany15b} examining the coherence in the results obtained for the extinction and distances. We analyzed our results in a similar way and conclude that the possible Cepheids are most likely to be type II from the background fields. These are old population stars lying behind the Galactic bulge. \\

\begin{table*}[!ht]
\small
\caption{Observed and derived parameters for the 5 Cepheid candidates of the two possible types}
\label{t:cep}
\centering
{\setlength{\tabcolsep}{5pt}
\begin{tabular}{lcccccccc}
\hline \hline
ID & $<K_s>$ & $<H-K_s>$ &  $E(H-K_s)_I$ & $(A_{K_s})_I$ & Dist$_I$ [pc] & $E(H-K_s)_{II}$ & $(A_{K_s})_{II}$ & Dist$_{II}$ [pc] \\
\hline
V11-Cls1 & 13.770 & 0.475 & 0.433 & 0.708 & 47133 & 0.384 & 0.627 & 18307  \\ 
V21-Cls1 & 12.991 & 0.369 & 0.330 & 0.540 & 31924 & 0.283 & 0.462 & 12738 \\ 
V38-Cls1 & 13.113 & 0.611 & 0.565 & 0.924 & 36835 & 0.512 & 0.837 & 13767 \\ 
V11-Cls2 & 13.682 & 0.608 & 0.568 & 0.928 & 38447 & 0.520 & 0.849 & 15164 \\ 
V58-Cls2 & 13.973 & 0.467 & 0.428 & 0.700 & 46803 & 0.381 & 0.621 & 18657 \\ 
\hline
\end{tabular}}
\end{table*}

\subsection{Eclipsing binaries analysis}

From the whole sample of binary system candidates, we selected the most probable ones and determined the physical parameters of the eclipsing components. We first used the Wilson \& Devinney (WD) code \citep{wd}, which operates in two steps while fitting the light curves, i.e., the LC (a subjective iteration) and the DC (an objective iteration) processes. The LC procedure is based on parameters previously determined from theory or observation \citep{w94a,w94b,w01,w06}. On the other hand, the DC is the differential calculus which aims at better determining the geometrical and physical parameters of the systems by reducing possible associated errors. Light curves were first analyzed by the LC procedure using known parameters, such as the period. The output file was then used as input for the DC procedure. The effectiveness of the WD code is not particularly favorable since it is time consuming so, as we intend to study many objects, we searched for another method. Currently, there are several graphical user interface programs that can be used for a scientific study of these objects. Among them, we tested the PHOEBE (PHysics Of Eclipsing BinariEs) program. This is a tool for modeling eclipsing binary stars based on photometric and spectroscopic (radial velocity) data, which is also based on the WD code. PHOEBE can determine the parameters associated with the physical and geometrical conditions of the system, like in the WD code.It can also deal with the parameters of the binary components to which we have access once the values associated with the obtained light-curves were determined.  \\ 

\begin{figure*}[!ht]
\centering
\includegraphics[width=\hsize]{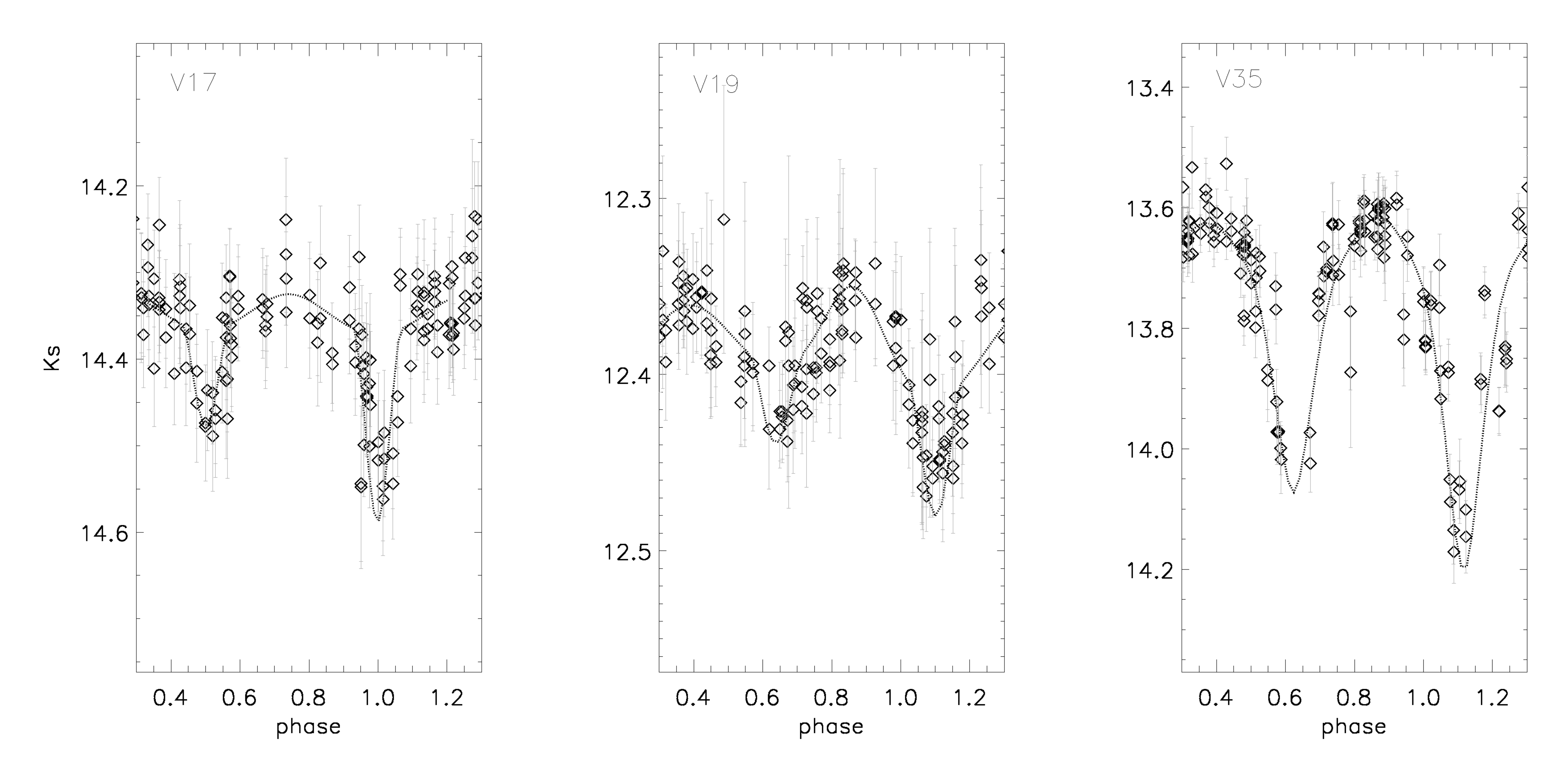}
\caption{Light curves of the best binaries found in the field of Antalova\,1. Solid lines represent the best fittings of the observed light curves for the eclipsing binary stars. Diamonds stand for the observed $<Ks>$ magnitudes taken from the VVV Survey.}
\label{f:cl1_bin}
\end{figure*}

\begin{figure*}[!ht]
\centering
\includegraphics[width=\hsize]{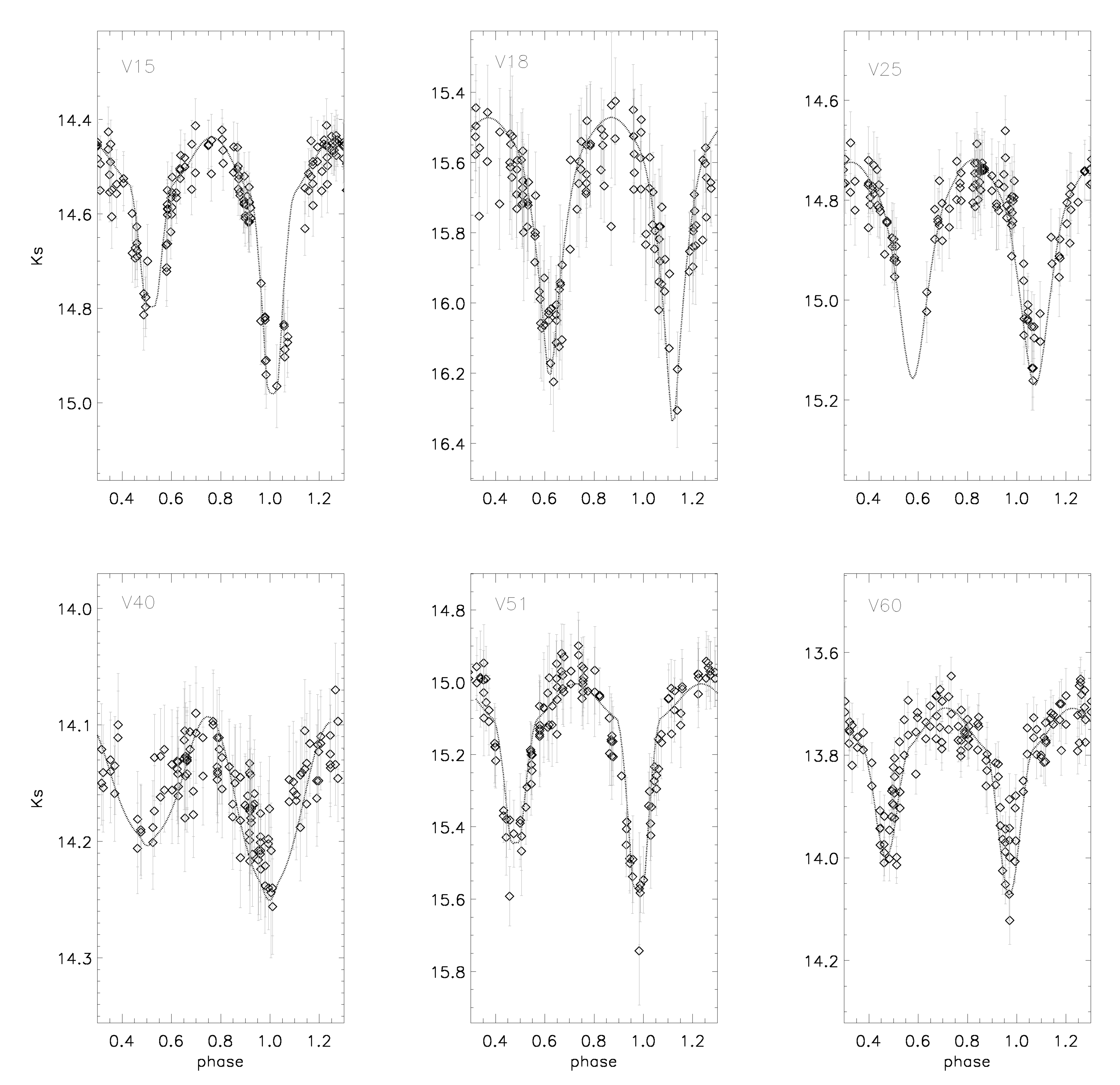}
\caption{Light curves of the best binaries found in the field of ASCC\,90. Lines and symbols represent the same as in Figure \ref{f:cl1_bin}.}
\label{f:cl2_bin}
\end{figure*}

\begin{figure*}[!ht]
\centering
\includegraphics[width=\hsize]{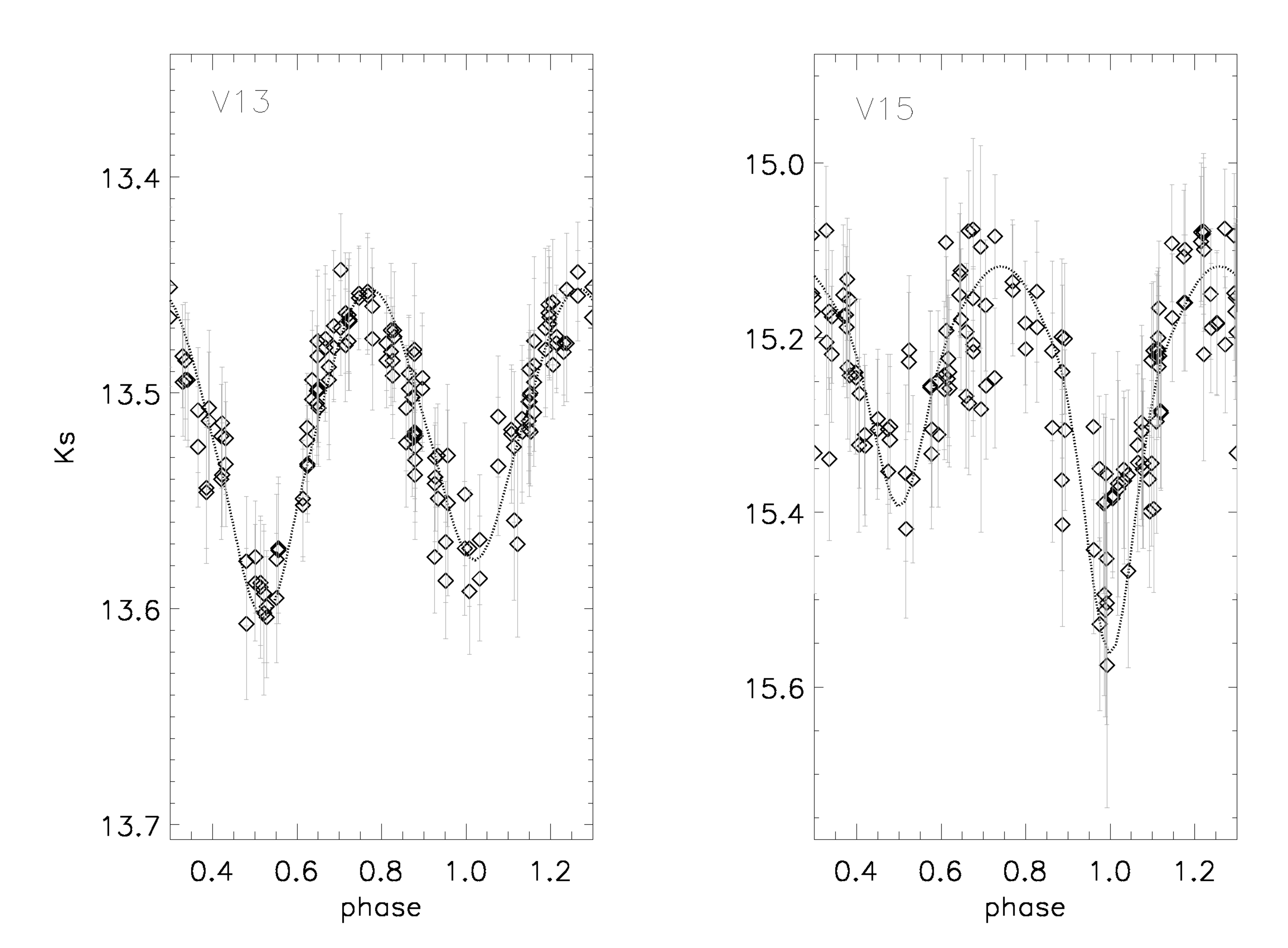}
\caption{Light curves of the best binaries found in the field of ESO\,393-15. Lines and symbols represent the same as in Figure \ref{f:cl1_bin}.}
\label{f:cl3_bin}
\end{figure*}

We used both WD and PHOEBE codes for our analysis and probed different classification modes. Taking into account the shape of the light curves, we conclude that the systems involved are detached, semi-detached or close contact binary systems. The best fittings of the corresponding light curves are shown in Figures \ref{f:cl1_bin} - \ref{f:cl3_bin}. Table \ref{t:bin} shows the resulting parameters that could be obtained from the current analysis. $M_2/M_1$ represents the mass ratio between the components. We would like to point out that, in general terms, the studied systems exhibit orbits that are almost circular, their eccentricities being about $5\times10^{-4}$. The stars involved seem to have solar or higher than solar surface temperatures. \\ 

\begin{table*}[!ht]
\small
\caption{Model and fixed parameters for the selected eclipsing binary stars.}
\label{t:bin}
\centering
{
\begin{tabular}{lccccccc}
\hline \hline
Eclipsing & Object Type & Period &  Inclination &  Eccentricity & $T_1$ & $T_2$ & $M_2/M_1$ \\
Binary &    & [days] & [$^{\circ}$] ($\pm$1.1)& ($\pm$0.006) & [K] & [K] & ($\pm$0.07) \\
\hline
V17-Cls1 & Detached & 1.7956 & 72.9 & 0.004 & 2900$\pm$80 & 2420$\pm$70 & 0.97 \\ 
V19-Cls1 & Detached & 38.0 & 75 & 0.000 & 8400$\pm$400 & 5000$\pm$200 & 0.69 \\ 
V35-Cls1 & Double-contact & 0.5163 & 77 & 0.009 & 3200$\pm$200 & 2900$\pm$200 & 0.80 \\ 
V15-Cls2 & Semi-detached & 0.5679 & 89 & 0.010 & 8000$\pm$400 & 5000$\pm$500 & 1.77 \\ 
V18-Cls2 & Double-contact & 0.3108 & 90 & 0.001 & 5000$\pm$200 & 4500$\pm$400 & 0.80 \\ 
V25-Cls2 & Double-contact & 0.7699 & 73.5 & 0.005 & 6000$\pm$70 & 5800$\pm$100 & 1.41 \\ 
V40-Cls2 & Semi-detached & 1.1193 & 63.5 & 0.005 & 7000$\pm$1000 & 3500$\pm$900 & 1.40 \\ 
V51-Cls2 & Semi-detached & 0.5435 & 90 & 0.000 & 6000$\pm$200 & 8000$\pm$200 & 1.80 \\ 
V60-Cls2 & Semi-detached & 5.3298 & 68 & 0.000 & 7000$\pm$1000 & 5600$\pm$600 & 0.79 \\ 
V13-Cls3 & Double-contact & 9.9522 & 50 & 0.000 & 3340$\pm$60 & 2950$\pm$50 & 0.80  \\ 
V15-Cls3 & Double-contact & 0.5980 & 68 & 0.000 & 4800$\pm$80 & 3500$\pm$30 & 0.90  \\ 
\hline
\end{tabular}}
\end{table*}

Since the cross-match is purely positional, most of the newly discovered variable stars are not cluster members but part of the bulge population projected onto the clusters' fields, such as RR Lyrae stars and the Cepheids found. According to the CMDs, we can obtain possible candidates for the clusters' members. In ASCC\,90 we found 43 possible EBs. However, only 8 of these binaries might be cluster members. In Antalova\,1, from 18 new found EBs, only 6 candidates could be cluster members and in ESO\,393-15,  from 18 BEs found in its region, only 3 of them might be cluster member candidates. The rest of the EBs found in ESO\,393-15 are probably background objects because they appear to have much redder colors than the corresponding cluster isochrone. Besides the detected variability and deep photometry, the good spatial resolution of the VVV Survey also allows for proper motion studies to be carried out in the future when a longer baseline becomes available. A further spectroscopic study is planned to obtain radial velocities which will help us to confirm or deny the physical association of the variables found to the corresponding clusters. The search in both cases, extended and compact OCs, present benefits as well as disadvantages. While the probabilities of finding variable stars in extended clusters are higher, their decontamination processes are more complicated, and the probabilities of containing field stars existing in the cluster regions are also higher. Our project intends to find the possible variables belonging to the host clusters, as well as to complete the search in the NIR bands and characterize the field variable stars. In our view, determining cluster membership is a difficult issue, contamination by background variable stars is a major problem and we believe that most of the variables found in the fields of the OCs studied here appear not to be associated with such OCs. A definite way to secure membership would be the use of kinematic data. Unfortunately, the radial velocities of these three clusters are largely unknown and quite telescope time consuming to obtain. However, their proper motions may be measured instead using the same VVV data. The accuracy of the proper motions measured by VVV astrometry has been estimated to be $\sim$ 2 mas/yr \citealt{libralato}. Such precision is not enough to secure cluster membership. We need  to increase the time baseline for some more years in order to obtain good and reliable (at least statistically) proper motions for the OCs of our study. \\

\section{Conclusions}
	
We have presented a new search for variable stars in the fields of three studied OCs using the NIR database of the VVV Survey. We have defined the procedures for variable stars search in VVV OCs. In this final section, a brief summary of our findings is presented.  \\

\begin{itemize}
\item We found 5 new Cepheids in the cluster fields, none of which appear to be cluster members. A large number of background RR Lyrae stars in the fields of the three studied clusters have also been found.\\
\item We identified a large number of EBs in the cluster fields. A total of 17, 42, and 8 EBs were found in the fields of  Antalova\,1, ASCC\,90 and ESO\,393-15, respectively. Based on NIR CMDs, we recognized only 8, 6, and 6 of such EBs as probable members of Antalova\,1, ASCC\,90 and ESO\,393-15, respectively. We obtained fundamental parameters for some selected EBs. They were classified as detached, semi-detached and/or double-contact binaries, with low eccentricities and high inclinations. Future spectroscopic follow-up of some of these binary systems may help to confirm cluster membership and also to obtain heliocentric distances to the clusters.
\item Our experience shows that for clusters projected onto highly reddened and obscured regions of the bulge, it is very hard to decontaminate the CMDs. Another major problem is to discriminate cluster members from field stars.
\item Lastly, it has not been possible to determine the variable type of some objects. That is mainly due to their lack of periodicity and to our inability to phase them using the current data. Therefore, they remain unclassified.
\end{itemize}

\section*{Acknowledgements}

We gratefully acknowledge the use of data from the ESO Public Survey program 179.B-2002 taken with the VISTA 4.1\,m telescope and data products from the Cambridge Astronomical Survey Unit. Support for TP, DM, ID and JAG is provided by the Ministry of Economy, Development, and Tourism's Millennium Science Initiative through grant IC120009, awarded to the Millennium Institute of Astrophysics, MAS. TP, JJC and LVG acknowledge financial support from the Argentinian institutions FONCYT, CONICET and SECYT (Universidad Nacional de C\'ordoba). DM is also supported by the Center for Astrophysics and Associated Technologies PFB-06, and Fondecyt Project No. 1130196. JAG acknowledges support from the FIC-R Fund, allocated to the project 30321072, by CONICYT's PCI program through grant DPI20140066 and from Fondecyt Iniciaci\'on 11150916. This research has made use of the SIMBAD database, operated at CDS, Strasbourg, France; also the SAO/NASA Astrophysics data (ADS).

\section{References}
\bibliographystyle{elsarticle-harv}
\bibliography{palma}

\end{document}